\documentclass[nofootinbib]{revtex4}
\pdfoutput=1  
\usepackage[latin1]{inputenc}
\usepackage{ae,aecompl}
\usepackage{amsmath,amssymb,mathrsfs}
\usepackage{graphicx}
\usepackage[]{units}
\usepackage[normalem]{ulem}
\usepackage{slashed} 
\usepackage[usenames,dvipsnames]{color} 
\usepackage{hyperref}
\hypersetup{
   colorlinks=true,       
    linkcolor=Blue,          
    citecolor=Plum,        
    filecolor=magenta,      
    urlcolor=YellowOrange           
}

\providecommand{\cM}{\mathcal{M}}

\providecommand{\msb}{\overline{\mathrm{MS}}}

\providecommand{\dm}[1][]{\delta m^2_{#1}}
\providecommand{\la}[1]{\lambda_{#1}}
\providecommand{\bla}[1]{\bar{\lambda}_{#1}}
\providecommand{\tv}{\tilde{v}}
\providecommand{\bv}{\bar{v}}
\providecommand{\hm}{\hat{m}}
\providecommand{\tp}{{\mss{\mathsf{T}}}}

\providecommand{\zb}[1][]{\overline{\mathrm{ZB}}_{#1}}
\providecommand{\zp}[1][]{\overline{\mathrm{ZP}}_{#1}}

\providecommand{\bvphi}{\bar{\varphi}}

\DeclareMathOperator{\sign}{sign}

\providecommand{\xlink}[1]
  {\href{http://arxiv.org/abs/#1}{#1}}
\providecommand{\eqali}[1]{\begin{equation}\begin{aligned} #1
    \end{aligned}\end{equation}}
\providecommand{\eq}[1]{\begin{equation} #1 \end{equation}}
\providecommand{\subeqali}[2][]{\begin{subequations}\label{#1}\begin{align}
#2    \end{align}\end{subequations}}
\providecommand{\ml}[1]{\mbox{\large $#1$}}
\providecommand{\mss}[1]{\mbox{\scriptsize $#1$}}
\providecommand{\ums}[2][1]{\ml{\tfrac{#1}{#2}}}
\providecommand{\aver}[1]{\langle #1 \rangle}
\providecommand{\mtrx}[1]{\begin{pmatrix} #1 \end{pmatrix}}
\DeclareMathOperator{\re}{\mathrm{Re}}
\DeclareMathOperator{\im}{\mathrm{Im}}
\DeclareMathOperator{\diag}{\mathrm{diag}}
\usepackage{mathtools}
\DeclarePairedDelimiter\inner{\langle}{\rangle}
\providecommand{\ZZ}{\mathbb{Z}}
\providecommand{\hs}[1]{\hspace{#1}}
\DeclareMathOperator{\tr}{\mathrm{tr}}

\usepackage{dsfont}

\newcommand{\Mh}{h}
\newcommand{\MH}{H}
\newcommand{\MA}{A}
\newcommand{\MG}{G}
\newcommand{\MHpm}{H^{\pm}}
\newcommand{\MGpm}{G^{\pm}}

\begin{document}
\title{
One-loop considerations for coexisting vacua in the CP conserving 2HDM
}
\author{A.~L.~Cherchiglia}
\email{adriano.cherchiglia@ufabc.edu.br}
\author{C.~C.~Nishi}
\email{celso.nishi@ufabc.edu.br}
\affiliation{
Centro de Matem\'{a}tica, Computa\c{c}\~{a}o e Cogni\c{c}\~{a}o\\
Universidade Federal do ABC - UFABC, 09.210-170,
Santo Andr\'{e}, SP, Brazil
}

\begin{abstract}
The Two-Higgs-Doublet model (2HDM) is a simple and viable extension of the Standard Model with
a scalar potential complex enough that two minima may coexist.
In this work we investigate if the procedure to identify our vacuum as the global minimum 
by tree-level formulas carries over to the one-loop corrected potential.
In the CP conserving case, we identify two distinct types of coexisting minima --- the regular
ones (moderate $\tan\beta$) and the non-regular ones (small or large $\tan\beta$) --- and conclude that the tree level 
expectation
fails only for the non-regular type of coexisting minima.
For the regular type, the sign of $m^2_{12}$ already precisely
indicates which minima is the global one, even at one-loop.
 \end{abstract}
\maketitle
\section{Introduction}
\label{sec:intro}

After the discovery of the Higgs boson of $125\,\unit{GeV}$ mass in 
2012\,\cite{higgs},
all the pieces of the SM were firmly established.
The Standard Model (SM) however is far from being a complete theory at the smallest physical scales 
as, to name a few shortcomings, it does not contain a Dark Matter candidate and cannot explain the matter and antimatter asymmetry of the Universe\,\cite{wmap}.

Among the various proposals,
considering an extended Higgs sector is one of the simplest modifications that we 
can implement in the SM without extending the hidden fundamental forces of nature
and also passing the many stringent collider tests at the LHC.
In this work we consider one of such extensions namely the Two-Higgs-doublet model 
(2HDM) in which just another scalar doublet is added.
This model features five physical Higgs bosons, three neutral and two charged, 
instead of just one in the SM.
It has been extensively studied in the literature (see e.g.\ Ref.\,\cite{branco.rev} for a review) partly because more fundamental theories, for instance 
the MSSM\;\cite{SUSY}, require a similar extended scalar sector.
The model also allows the possibility of other sources of CP violation\,\cite{lee},
a feature that gets even richer when more doublet copies are added\,\cite{nhdm:cp}.
Finally, a more complex scalar sector can generate a strong enough EW first-order phase 
transition\,\cite{2hdm:PT,2hdm:PT:recent}, a property that is lacking in the SM\,\cite{sm:PT} but is necessary to explain the matter-antimatter asymmetry of our universe.

Another feature that a more involved scalar potential encompasses is the possibility of different symmetry breaking patterns and the 2HDM is no exception\,\cite{deshpande.ma,charge.breaking,coexisting.min,ivanov:mink,pilaftsis}.
With additional Higgs doublets, this complexity increases 
substantially\,\cite{nhdm}.
In general, for a sufficiently complex potential, 
there may even exist sets of parameters for which many local minima coexist.
In this case, identifying which one is the global minimum 
might be a nontrivial task 
that involves solving a system of polynomial equations.
Usually, when one is sure that only \textit{one} minimum is present for a 
given choice of parameters, such a task is bypassed by trading some of the quadratic 
parameters in favor of the vevs and ensuring the extremum is a local minimum.
In the 2HDM, that assumption does not hold in general and 
for some parameter ranges it is possible that up to two minima coexist for the same
potential\,\cite{ivanov:mink}.
So a worrisome possibility arises: we may be living in a 
metastable
vacuum with the possibility to tunnel to the global minimum.
That situation was described in Ref.\,\cite{panic.vacuum} as our vacuum being a 
panic vacuum.
One way of testing that situation is explicitly calculating the depth of the second 
minimum and comparing it to the depth of the first.
However, finding the location of the minima explicitly may be a difficult
or, at least, computationally intensive task.
Fortunately, in the same work, the authors developed a method capable of 
distinguishing if our vacuum is a panic vacuum by calculating a 
\textit{discriminant} that depends only on the position of \textit{our vacuum}
(see Ref.\,\cite{discriminant:general} for a more general test).
Although many of the possible scenarios
 with coexisting minima are not favored by current
LHC data\,\cite{panic.vacuum},
we are interested here in studying if the simple use of this discriminant can be carried over to
the one-loop corrected effective potential.
Already for the inert doublet model\,\cite{deshpande.ma} it was found that the potential difference of the coexisting 
minima can change sign when one-loop corrections are taken into account\,\cite{pedro}. 
Therefore, the present work aims to verify the validity of such conclusions for a general
CP conserving 2HDM with softly broken $\ZZ_2$.
We focus on the case of two coexisting normal vacua and study the predictive power of tree-level formulas for the depth of the potential when one-loop corrections are considered.

The outline of the paper is as follows: In Sec.\,\ref{sec:coexist} we review the 
properties of
the 2HDM with softly broken $\ZZ_2$ at tree-level focusing on the possibility of
two coexisting minima.
Some results can not be found in previous literature.
In Sec.\,\ref{sec:eff} we review the form of the one-loop effective potential for 
our
case while Sec.\,\ref{sec:min} explains our procedure for ensuring that our vacuum
has the correct vacuum expectation value.
The procedure to compute and fix the pole mass of the SM Higgs boson to its 
experimental value is explained in Sec.\,\ref{sec:pole}.
The steps we performed to generate the numerical samples are listed in 
Sec.\,\ref{sec:numerical}
and the resulting analysis is shown in Sec.\,\ref{sec:results}.
Finally, the conclusions can be found in Sec.\,\ref{sec:conclusions}.

\section{Coexisting normal vacua at tree-level}
\label{sec:coexist}

The general 2HDM potential at tree-level is
\eqali{
\label{pot:tree}
V_0&=m^2_{11}|\phi_1|^2+m^2_{22}|\phi_2|^2
-\big[m^2_{12}\phi_1^\dag\phi_2+h.c.\big]
+\ums{2}\lambda_1|\phi_1|^4
+\ums{2}\lambda_2|\phi_2|^4
\cr
&\quad +\ \lambda_3|\phi_1|^2|\phi_2|^2+\lambda_4|\phi_1^\dag\phi_2|^2+
\Big\{
\ums{2}\lambda_5(\phi_1^\dag\phi_2)^2+
\big[\lambda_6|\phi_1|^2+\lambda_7|\phi_2|^2\big]\phi_1^\dag\phi_2+h.c.
\Big\}
\,.
}
We will be considering the real softly broken $\ZZ_2$ symmetric case where 
$m^2_{12}$ and $\la5$ are real while $\lambda_6=\lambda_7=0$.

We will also focus on CP conserving vacua where the vacuum expectation values are 
real
and we employ the parametrization
\eq{
\label{vevs}
\aver{\phi_1}=\frac{1}{\sqrt{2}}\mtrx{0\cr v_1}\,,\quad
\aver{\phi_2}=\frac{1}{\sqrt{2}}\mtrx{0\cr v_2}\,.
}
These vevs can be further parametrized by modulus and angle as
\eq{
(v_1,v_2)=v\,(c_\beta,s_\beta)\,,
}
where $v=246\,\unit{GeV}$ for our vacuum and we use the shorthands 
$c_\beta\equiv\cos\beta$, 
$s_\beta\equiv\sin\beta$.
We call this type of vacuum a \textit{normal vacuum} and we denote our 
vacuum by NV\,\cite{panic.vacuum} with $v_1>0$ and $v_2>0$.

By ensuring the existence of one normal vacuum (our vacuum), a scalar potential with 
fixed 
parameters cannot simultaneously have another minimum of a different type, namely a 
charge breaking vacuum or a spontaneously CP breaking vacuum\,\cite{charge.breaking}.
Just another coexisting normal vacuum NV$'$ with vevs $(v_1',v_2')$ may exist and
this is the only case 
where two minima can coexist in the 2HDM potential at tree level\,\cite{ivanov:mink}:
\textit{only two minima with the same residual symmetry may coexist}.
When the coexisting minima exist, we define the potential difference as
\eq{
\label{depth.diff}
\Delta V\equiv V_{\rm NV'}-V_{\rm NV},
}
so that $\Delta V>0$ indicates that our vacuum is the global minimum.
We use this convention for the one-loop potential as well.

To describe the situation of two coexisting normal vacua in more detail, we can 
write the extremum equations for nonzero $v_1$ and $v_2$:
\eqali{
\label{extremum:tree}
\frac{\partial V_0}{\partial v_1}=
v_1(m^2_{11}+\ums{2}\lambda_1v_1^2+\ums{2}\lambda_{345} v_2^2)
-m^2_{12}v_2
&=0
\,,
\cr
\frac{\partial V_0}{\partial v_2}=
v_2(m^2_{22}+\ums{2}\lambda_2v_2^2+\ums{2}\lambda_{345}v_1^2) 
-m^2_{12}v_1
&=0
\,.
}
We employ the usual shorthand $\lambda_{345}\equiv \lambda_3+\lambda_4+\lambda_5$.

We will see that there are two types of coexisting normal vacua depending on the
$\ZZ_2$ symmetric limit.
The complete solutions of Eq.\,\eqref{extremum:tree} for $m^2_{12}=0$ can be easily found: there are
two degenerate extrema that spontaneously break $\ZZ_2$ --- ZB$_{+}$ and ZB$_{-}$ ---
and two extrema that preserve $\ZZ_2$ --- ZP$_1$ and ZP$_2$; the latter are often denoted as
inert or inert-like vacuum (see Ref.\,\cite{pedro} and references therein).
Only one of the pairs ZP$_{1,2}$ or ZB$_{\pm}$ may coexist as minima.%
\footnote{Note that one of ZP$_{1,2}$ may not be a minimum whereas ZB$_{\pm}$ are always
degenerate.}
They are characterized by $(v_1,v_2)$ of the form
\eqali{
\text{ZB}_+:\quad (\bv_1,\bv_2)\,,&
\quad
\text{ZB}_-:\quad (\bv_1,-\bv_2)\,,
\cr
\text{ZP}_1:\quad (\tv_1,0)\,,&
\quad
\text{ZP}_2:\quad (0,\tv_2)\,,
}
where we adopt the convention that all $\bv_1,\bv_2,\tv_1,\tv_2$ are positive.
The specific values of the vevs are given by
\eq{
\label{X:Z2}
\mtrx{\bv^2_1\cr \bv^2_2}=
\mtrx{\la1 & \la{345}\cr \la{345} & \la2 }^{-1}
\mtrx{-2m^2_{11}\cr -2m^2_{22}}
\,,
}
for the $\ZZ_2$ breaking extrema\footnote{%
As long as the solutions for $v^2_i$ give positive solutions.}
and
\eq{
\tv_1^2=\frac{-2m^2_{11}}{\la1}>0\,,\quad
\tv_2^2=\frac{-2m^2_{22}}{\la2}>0\,,
}
for the $\ZZ_2$ preserving minima.
The two extrema ZB$_\pm$ are indeed connected by the spontaneously
broken $\ZZ_2$ symmetry: $\phi_2\to -\phi_2$.
We note that simultaneous sign flips of both $v_1,v_2$ is a gauge symmetry and do 
not count as a degeneracy. Hence we adopt the convention that $v_1>0$ while $v_2$
can attain both signs so that we only
analyze the first and fourth quadrant in the $(v_1,v_2)$ plane.

As the $-m^2_{12}v_1v_2$ term is continuously turned on, the $\ZZ_2$ symmetry is soft but
explicitly broken with a negative (positive) contribution in the first (fourth) quadrant 
when $m^2_{12}>0$. 
The opposite is true for negative $m^2_{12}$.
The effect of adding the $m^2_{12}$ term is different for the two types of
coexisting minima which we denote by $\zb[\pm]$ and $\zp[1,2]$ from their $m^2_{12}\to 0$ limit.
We also denote the $\zb[\pm]$ minima as \textit{regular} and $\zp[1,2]$ as \textit{non-regular}
simply because it is much more probable to generate models with the former pair than the latter
for generic values of $\tan\beta$ and other parameters.

The two degenerate spontaneously breaking minima $\mathrm{ZB}_{\pm}:(\bar{v}_1,\pm\bar{v}_2)$ deviate to $\zb[+]:(v_1,v_2)$ and $\zb[-]:(v_1',v_2')$, respectively, and the
degenerate potential depth,
\eq{
V_0(\bar v_1,\bar v_2)=-\ums{2}\mtrx{m^2_{11} & m^2_{22}}
\mtrx{\la1 &\la{345}\cr \la{345} &\la2}^{-1}\mtrx{m^2_{11}\cr m^2_{22}}
\,,
}
also deviates differently lifting the degeneracy.
In first approximation in small $m^2_{12}$ and in the deviation of the vevs, the potential depths change respectively by the amount
$
\delta V_{\pm}\approx \mp m^2_{12}\bar{v}_1\bar{v}_2\,,
$
so that the depth difference of the two minima is
\eq{
\label{Delta.V:ZB}
\Delta V_{\zb}\equiv V_{\zb[-]}-V_{\zb[+]}=
\delta V_{-}-\delta V_{+}\approx 2m^2_{12}\bar{v}_1\bar{v}_2\,.
}
See appendix \ref{ap:deviation} for the general formula.
As we defined our vacuum to be $\zb[+]$, we see that indeed it gets \textit{deeper} as
$m^2_{12}$ \textit{increases} from zero while the non-standard vacuum $\zb[-]$ is 
pushed up.

The deviation for the $\ZZ_2$ preserving minima are different: the first order perturbation
to the potential value vanishes.
We need the deviations in the locations of the minima $\zp[1,2]$ which, in first order, read
\eq{
\zp[1]: (\tv_1,\delta v_2)\,,
\quad
\zp[2]: (\delta v_1,\tv_2)\,,
}
where
\eq{
\delta v_2\approx\frac{m^2_{12}}{m^2_{H_2}}\tv_1\,,
\quad
\delta v_1\approx\frac{m^2_{12}}{m^2_{H_1}}\tv_2\,,
}
with $m^2_{H_i}=m^2_{ii}+\ums{2}\la{345}\tv_j^2$, $(ij)=(2,1)$ or $(1,2)$, is the second derivative in the $v_i$ direction around ZP$_j$.
When $m^2_{12}>0$, the deviations $\delta v_i$ are positive and the two minima enter 
the first quadrant. Otherwise they move to the fourth quadrant.
The potential depth then changes from the $\ZZ_2$ limit
\eq{
V_0(\tv_1,0)=-\frac{m^4_{11}}{2\la1}\,,\quad
V_0(0,\tv_2)=-\frac{m^4_{22}}{2\la2}\,,
}
by
\eq{
\delta V_0|_{\zp[1]}\approx-\ums{2}m^2_{H_2}(\delta v_2)^2\,,
\quad
\delta V_0|_{\zp[2]}\approx-\ums{2}m^2_{H_1}(\delta v_1)^2\,,
}
respectively.
The depth difference of the coexisting $\zp[1,2]$ is
\eq{
\Delta V_{\zp}=V_0|_{\zp[1]}-V_0|_{\zp[2]}\approx
-\Big(\frac{m^4_{11}}{2\la1}-\frac{m^4_{22}}{2\la2}\Big)
-\ums{2}m_{12}^4\Big(\frac{m^2_{11}}{\la1 m^2_{H_2}}-\frac{m^2_{22}}{\la2 m^2_{H_1}}\Big)
\,.
}
The first term corresponds to the usual difference between the two inert-like vacua.
We conventionally consider $\zp[2]$ to be our vacuum NV corresponding to large $\tan\beta$.

\begin{figure}[h]
\includegraphics[scale=0.405]{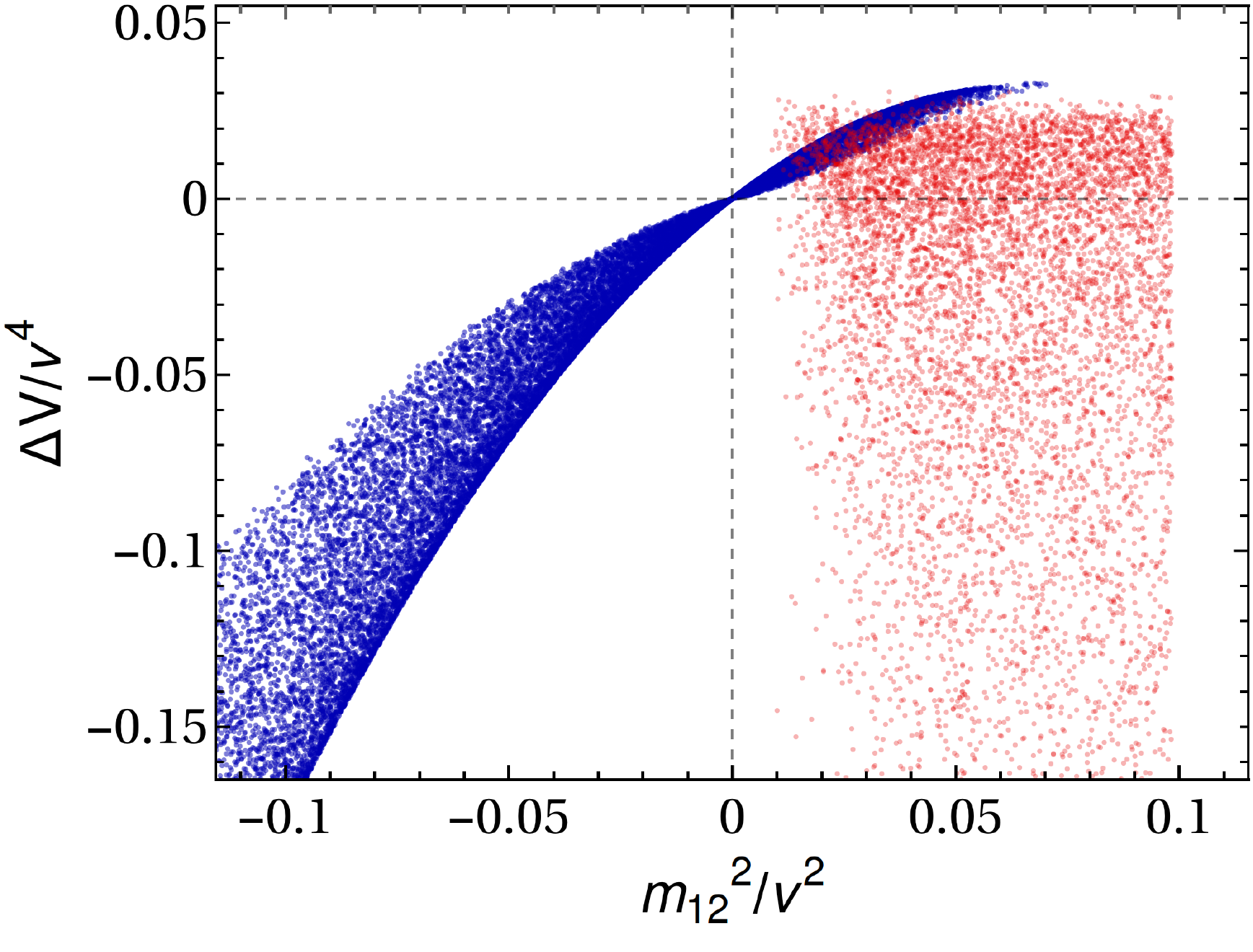}
\includegraphics[scale=0.4]{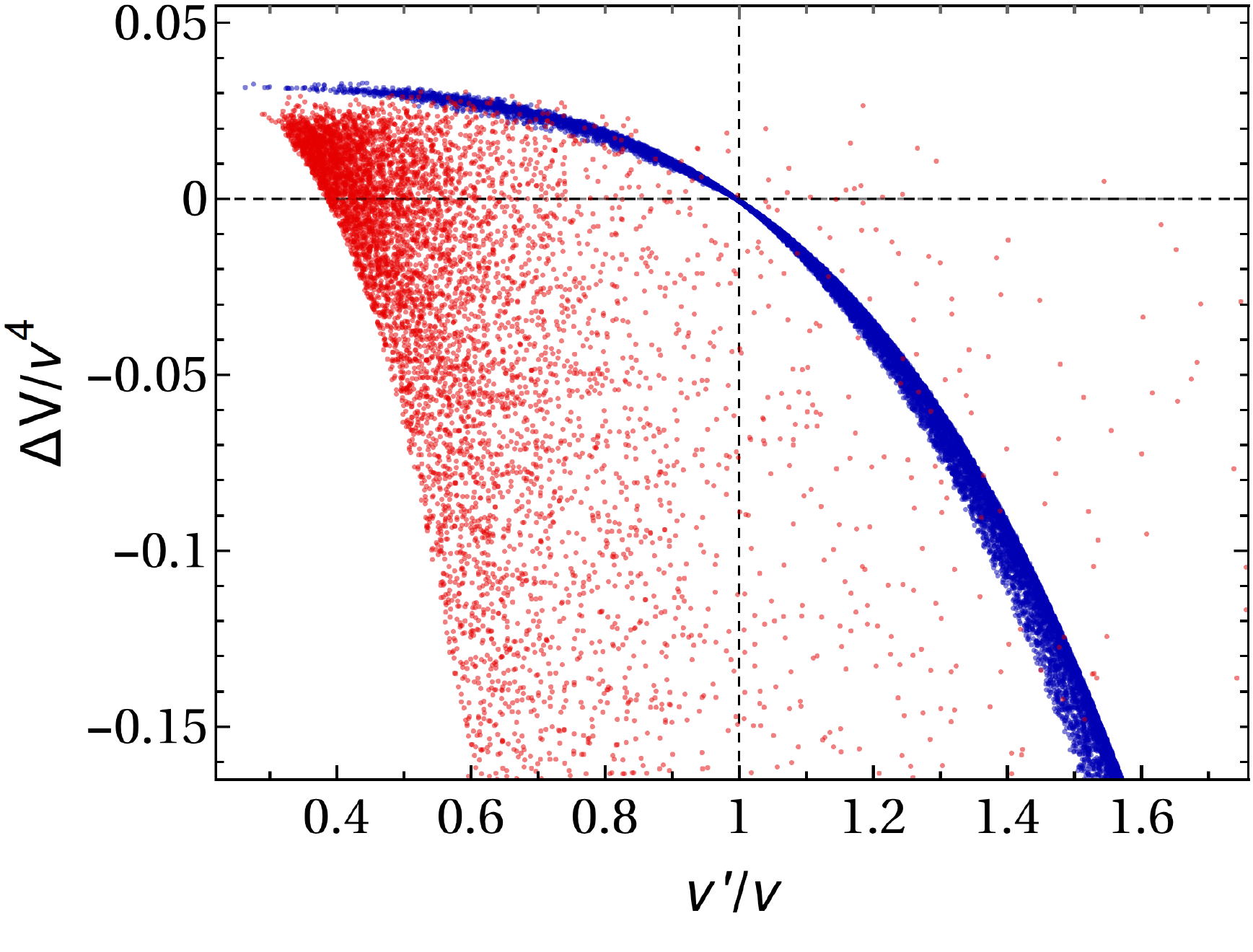}
\caption{
\textbf{Left:}
Potential difference \eqref{depth.diff} at the two coexisting minima against $m^2_{12}$, both normalized by $v=246\,\unit{GeV}$.
\textbf{Right:}
The same depth difference against the ratio of non-standard vev (NV$'$) to the standard vev (NV).
}
\label{fig.1}
\end{figure}

The behaviors described above can be clearly seen in the left panel of 
Fig.\,\ref{fig.1} (blue points) for $\zb[\pm]$ where we show the
depth difference calculated exactly against $m^2_{12}$, both normalized by the 
appropriate
power of $v=246\,\unit{GeV}$ (NV).
Only potentials with two minima are selected and the free parameters are taken as 
\eq{
\{\tan\beta,\alpha,m^2_{H^+},m^2_{A},m^2_{H^0},m^2_{h},m^2_{12}\}\,,
}
with $m_h=125\,\unit{GeV}$, $1\le\tan\beta\le 50$, $\{m_{H^+},m_A,m_H\}$ ranging from 
90\;GeV to 1\;TeV ($m_H>m_h$),
$-20{,}000\,\unit{GeV}^2\le m^2_{12}\le 6000\,\unit{GeV}^2$
and $\alpha$ is constrained
near alignment, $-0.1\le\cos(\beta-\alpha)\le 0.1$.
Simple bounded from below and perturbativity constraints are also imposed\,\cite{branco.rev}.
The blue points end around $m^2_{12}/v^2\sim 0.07$ because the nonstandard minimum 
gets pushed
up until the point where it disappears.
In contrast, the right panel shows the normalized depth difference with respect to
the ratio $v'/v$ of the values for NV$'$ and NV.
We can see for the blue points that the vacuum that lies deeper has a larger vacuum expectation
value.
The method we employed to calculate the location of NV$'$ is described in appendix \ref{ap:NV'}.

We confirm the approximation \eqref{Delta.V:ZB}: for $\zb[\pm]$ the sign of $m^2_{12}$ discriminates between our vacuum being the global minimum ($m^2_{12}>0$) or just a local metastable vacuum ($m^2_{12}<0$).
This behavior, however, does not apply for the points (red) that deviate from the inert-like vacua, $\zp[1,2]$, where the nonstandard vacuum may lie deeper despite $m^2_{12}$ being positive.
For the generic values of $t_\beta$ as used above the density of non-regular coexisting minima 
is very low so that only a handful of coexisting non-regular minima is obtained jointly with the regular points.
To generate a sufficient number of non-regular points we further 
produced another sample
(most of the red points) by restricting $20 \le t_\beta\le 50$ and positive $m^2_{12}$.

\begin{figure}[h]
\includegraphics[scale=0.4]{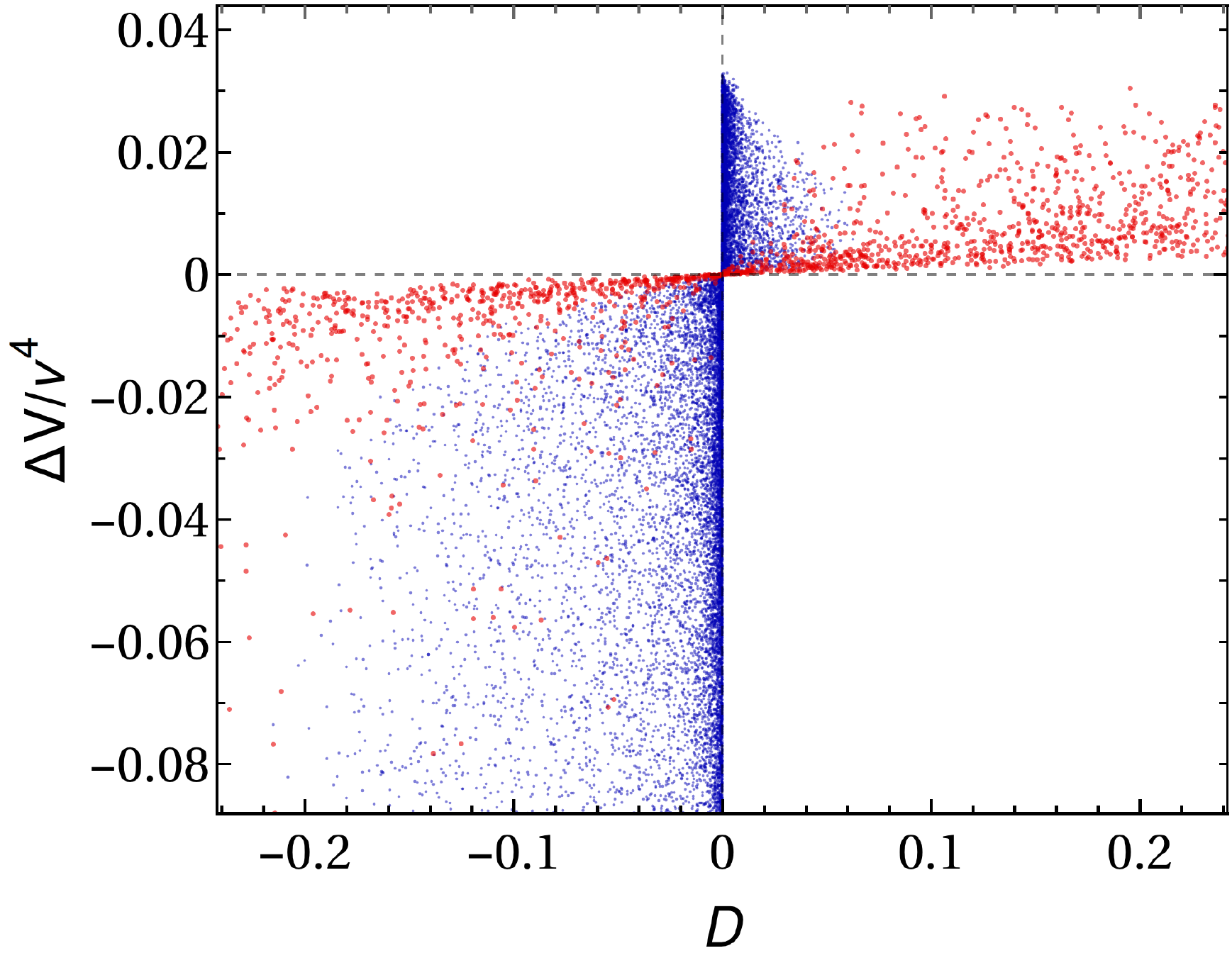}
\caption{
Normalized potential difference \eqref{depth.diff} at the two coexisting minima against the discriminant \eqref{discriminant}.
}
\label{fig.2}
\end{figure}

To accurately distinguish among the different cases, Ref.\,\cite{panic.vacuum} 
constructed a very useful discriminant $D$ that ensures that our vacuum is always 
the global minimum
if $D$ is positive.\footnote{For $D=0$ but $m^2_{12}\neq 0$ the discriminant is 
inconclusive.}
Since that discriminant was derived assuming that $v_1,v_2$ are both positive, 
we cannot apply it to NV$'$ when $v_2'<0$.
So we rederived the discriminant allowing the vevs to be negative with the result
\eq{
\label{discriminant}
D=\frac{1}{v^4}m^2_{12}(m^2_{11}-k^2m^2_{22})s_{2\beta}(t^2_\beta-k^2)\,,
}
where $k\equiv (\la1/\la2)^{1/4}$ and we have normalized to obtain a dimensionless
quantity.
This discriminant is useful because it can be obtained by using only
the angle $\beta$ calculated in one vacuum and cases with only one minimum are 
automatically taken into account.
The discriminating power of $D$ is shown in Fig.\,\ref{fig.2} where the depth 
difference is plotted against $D$ calculated using NV. Obviously we could have
calculated the discriminant for NV$'$, obtaining a $D'$ with sign opposite to $D$.
That implies that the quantity that depends on the vevs,  $s_{2\beta}(t^2_\beta-k^2)$,
must have opposite signs when calculated for NV and NV$'$.

Our main goal here is to analyze if the discriminant power of $m^2_{12}$ and $D$
carries over to the one-loop effective potential.

\section{Effective potential at one-loop}
\label{sec:eff}

We can now consider the effective potential
\eq{
\label{V0+V1}
V=V_0+V_{1l}\,,
}
with the one-loop contribution 
\eq{
\label{V1l}
V_{1l}=\frac{1}{64\pi^2}
\sum_k c_k 
M^4_k(\varphi_i)\Big(\ln\frac{M_k^2(\varphi_i)}{\mu^2}-\frac{3}{2}\Big)\,.
}
The masses $M^2_k(\varphi_i)$ correspond to the scalar-field-dependent 
eigenvalues of the tree-level mass matrices of all particles of the theory while 
$\mu$ is the renormalization scale.
We are already assuming a renormalization scheme with minimal subtraction ($\msb$) 
and, for the gauge sector, the Landau gauge and dimensional reduction (DRED),
following the scheme of Ref.\,\cite{martin}.
The parameters contained in $V_0$ are thus the renormalized parameters.
The integer coefficients $|c_k|$ count all the degrees of freedom for each 
particle $k$ including color, charge and spin, while the sign of $c_k$ is 
determined by its boson/fermion character:
positive for bosons and negative for fermions. 
For example, for the top quark we have $c_t=-3\times 2\times 2$ corresponding to its 3 colors, 2 
particle/antiparticle and 2 spin degrees of freedom.
We should note that the effective potential is generically a gauge dependent quantity
but its value at an  extremum is not\,\cite{nielsen}.

As we will focus on normal vacua, we can consider that the effective potential depends only
on the two real values $\varphi_1,\varphi_2$ in the real neutral directions%
\,\footnote{For a generic field dependence modulo gauge freedom, we would need two more real directions.}:
\eq{
\label{varphii}
\phi_1=\frac{1}{\sqrt{2}}\mtrx{0\cr \varphi_1}\,,\quad
\phi_2=\frac{1}{\sqrt{2}}\mtrx{0\cr \varphi_2}\,.
}
We reserve the symbols $v_1,v_2$ in Eq.\,\eqref{vevs} to values at a minimum. So 
the field-dependent gauge boson masses retain the same functional form as in the SM with 
$v^2=v_1^2+v_2^2$:
\eq{
M_W^2(\varphi_i)=\ums{4}g^2(\varphi_1^2+\varphi_2^2)\,,\quad
M_Z^2(\varphi_i)=\ums{4}(g^2+g'^2)(\varphi_1^2+\varphi_2^2)\,.
}
The fermion masses depend on the type of 2HDM we are considering.
We only consider models where FCNC are suppressed due to a $\ZZ_2$ symmetry.
We focus more on the type I model but our results are equally valid for all types II, X and Y
because of the dominance of the top Yukawa.
For the type I we have
\eq{
M_t(\varphi_i)=\frac{y_t}{\aver{s_\beta}}\frac{\varphi_2}{\sqrt{2}}\,,\quad
M_b(\varphi_i)=\frac{y_b}{\aver{s_\beta}}\frac{\varphi_2}{\sqrt{2}}\,,
}
where $y_{t,b}$ are the Yukawa couplings of the third family quarks normalized to the SM values and the enhancement factor $1/\aver{s_\beta}$ should be
considered as the fixed value at the NV minimum at one-loop. We emphasize that information with the bracket $\aver{~~}$.
For the type II, we should replace the $M_b$ dependence on $\varphi_2$ and $\aver{s_\beta}$
by $\varphi_1$ and $\aver{c_\beta}$ respectively.
We will see that, as usual, the top correction dominates the fermion loops and the 
difference between type I or type II is negligible for the one-loop corrections 
except for excessively large $\tan\beta$ which we do not consider.
It is also justified that we only consider the effects of the top and bottom quarks;
see Fig.\,\ref{fig.3} and comments in the text.

For the scalar contribution we need to calculate the eigenvalues of the matrix of second derivatives of
$V_0$ for \textit{generic} values of $\varphi_i$.
These mass matrices are shown in appendix \ref{ap:ddV} and their eigenvalues correspond to 
$M^2_S(\varphi_i)$ of the 8 scalars $S\in\{G^\pm,H^\pm,G^0,A^0,H^0,h^0\}$.
Due to charge and CP conservation, the mass matrices are still separated into three sectors: two charged scalars and its antiparticles, two CP odd scalars and two CP even scalars.
We emphasize that e.g.\ $M^2_{G^0}(\varphi_i)$ is nonvanishing at $\varphi_i$ away from any tree-level minimum.
It is the second derivative of the whole effective potential at one-loop 
that will vanish in the directions of the Goldstone modes.

\section{Parametrization and minimization at one-loop}
\label{sec:min}

We are interested in surveying the cases where the effective potential at one-loop \eqref{V0+V1}
continues to have two local minima, one of which should be our vacuum with $v=\sqrt{v^2_1+v^2_2}=246\,\unit{GeV}$.
The vevs $v_1,v_2$ no longer satisfy the tree-level minimization relations in \eqref{extremum:tree}
but should now minimize the whole effective potential $V_0+V_{1l}$.
We need a convenient parametrization to ensure that one minimum has the appropriate value of $v$.

To parametrize $V_0$, we will use as input the usual 8 quantities
\eq{
\label{input}
\{v_1,v_2,\alpha,m^2_{H^+},m^2_{A},m^2_{H^0},m^2_{h},m^2_{12}\},
}
where $v_i$ satisfy the minimization equations \eqref{extremum:tree}.
It is clear that these quantities define 
$V_0$ unambiguously by fixing the 8 parameters $\{m^2_{11},m^2_{22},m^2_{12},\la1,\la2,\la3,\la4,\la5\}$; see e.g.\ the first reference in \cite{panic.vacuum}.
When we add the one-loop contribution, it is clear that the true minimum will be shifted 
by a small amount from the position $(v_1,v_2)$ at tree-level.
Instead of correcting for that shift, we add the finite counterterms
\eq{
\delta V= \dm[11]|\phi_1|^2+\dm[22]|\phi_2|^2\,,
}
to the potential and \textit{adjust} the values of $\dm[ii]$ so that $v_1,v_2$ continue to be
a minimum at one-loop.\,%
\footnote{%
This procedure is equivalent to trading $m^2_{11},m^2_{22}$ in favor of 
$v_1,v_2$\,\cite{pedro}.
}
This means that the one-loop effective potential \eqref{V0+V1} is now rewritten as
\eq{
\label{V0+dV+V1l}
V=V_0+\delta V+V_{1l}\,.
}
We can see that $\dm[ii]\to 0$ and $V_{1l}\to 0$
in the limit where we turn off all couplings of the scalars to other particles including self-couplings.
For small couplings, it is also expected that the physical masses are close to the masses
$\{m^2_{H^+},m^2_{A},m^2_{H^0},m^2_{h}\}$ used as input.
It is possible to use a different renormalization scheme where all the masses and mixing angles at tree level are maintained at one-loop\,\cite{krause}.
Our scheme, however, avoids the need to deal with infrared divergences coming from 
the vanishing Goldstone masses\,\cite{cline,2hdm:PT:recent}.
This problem is more severe at higher loop orders\,\cite{martin.14}.

Now the minimization equations at one-loop can be separated into a tree-level part
\eq{
\frac{1}{\varphi_i}\frac{\partial V_0}{\partial \varphi_i}\bigg|_{\varphi_i=v_i}=0\,,~~i=1,2,
}
which leads to the tree-level equations written in \eqref{extremum:tree},
and a one-loop part that defines $\dm[ii]$ by
\eq{
\label{eq:deltam}
-\dm[ii]=\frac{1}{\varphi_i}\frac{\partial V_{1l}}{\partial \varphi_i}\bigg|_{\varphi_i=v_i}\,,~~i=1,2.
}
We can separate the derivative of $V_{1l}$ into its contribution from scalars (S),
vector bosons (V) and fermions (F):
\eqali{
\label{extremum:1-loop}
\frac{1}{\varphi_i}\frac{\partial V^S_{1l}}{\partial \varphi_i}\bigg|_{\varphi_i=v_i}
&=\frac{1}{32\pi^2}\times
\sum_{\begin{split}\mss{S=}&\,\mss{G^\pm,H^\pm,G^0,}\\[-2ex] 
&\,\mss{A^0,H^0,h^0}\end{split}}
\la{i S}\,M^2_S\bigg(\ln\frac{M^2_S}{\mu^2}-1\bigg)\,,
\cr
\frac{1}{\varphi_i}\frac{\partial V^V_{1l}}{\partial \varphi_i}\bigg|_{\varphi_i=v_i}
&=\frac{3}{32\pi^2}\bigg\{
\frac{g^2+g^{\prime 2}}{2}m^2_Z\bigg[\ln\frac{m^2_Z}{\mu^2}-1\bigg]
+g^2m^2_W\bigg[\ln\frac{m^2_W}{\mu^2}-1\bigg]
\bigg\}\,,
\cr
\frac{1}{\varphi_i}\frac{\partial V^F_{1l}}{\partial \varphi_i}\bigg|_{\varphi_i=v_i}
&= \frac{-4}{32\pi^2}\bigg\{\frac{3y_t^2}{\aver{s_\beta}^2}
m^2_t\bigg[\ln\frac{m^2_t}{\mu^2}-1\bigg]
+\frac{3y_b^2}{\aver{s_\beta}^2}
m ^2_b\bigg[\ln\frac{m^2_b}{\mu^2}-1\bigg]
\bigg\}\,,
}
for each $i=1,2$.
The dimensionless coefficients $\lambda_{i S}$ are given in appendix \ref{ap:dddV}
and we use lowercase letters for $m_{W,Z,t,b}$ because they correspond to the actual values
in the SM when we use our vacuum NV.
For charged particles the contribution of the antiparticle can be taken into account 
by doubling the contribution of the particle.
The fermion part corresponds to the type I model. For the type II model, we must replace
$\aver{s_\beta}\to \aver{c_\beta}$ in the couplings to the $b$ quark.
We can see in Fig.\,\ref{fig.3} that the contributions from scalars are large for
$\dm[11]$ and $\dm[22]$ while the top contribution is also large and negative for $\dm[22]$.
The contribution from the bottom quark is negligible for $\tan\beta\le 50$ and there is no appreciable difference between the type I or type II model. Thus for definiteness we consider
the type I model.
We also note that the scalar masses and coefficients depend on $\dm[11],\dm[22]$ and 
\eqref{eq:deltam} must be solved self-consistently.

The only remaining task is to write $M^2_S(v_i)$ in terms of the input parameters
\eqref{input}.
We note that $M^2_S(v_i)$ should be computed from the second derivatives of $V_0+\delta V$. But the part coming from $V_0$ at $\varphi_i=v_i$ corresponds to the usual masses at tree-level because $(v_1,v_2)$ still corresponds to a minimum of $V_0$.
Therefore, these matrices will have the generic form
\eq{
\label{shift:mass}
\cM(v_i)=\cM^{\rm tree}+\delta\cM\,.
}
Specifically, the mass matrices for the different sectors read
\eq{
\label{M:charged}
\cM_{\rm char}(v_i)=m^2_{H^+}\mtrx{s^2_\beta & -s_\beta c_\beta\cr -s_\beta c_\beta & 
c^2_\beta}
+\mtrx{\dm[11]&\cr&\dm[22]}
\,,
}
\eq{
\label{M:odd}
\cM_{\rm odd}(v_i)=m^2_A\mtrx{s^2_\beta & -s_\beta c_\beta\cr -s_\beta c_\beta & 
c^2_\beta}
+\mtrx{\dm[11]&\cr&\dm[22]}\,,
}
\eq{
\label{M:even}
\cM_{\rm even}(v_i)=
\left(
\begin{array}{cc}
 \la{1} v^2 c_\beta^2+m^2_{12} t_\beta  & -m^2_{12}+\la{345}v^2 s_\beta c_\beta  \\
 -m^2_{12}+\la{345}v^2 s_\beta c_\beta  & \la{2}v^2 s^2_\beta+m^2_{12}\, \cot_\beta 
\\
\end{array}
\right)
+\mtrx{\dm[11]&\cr&\dm[22]}\,,
}
where $m^2_{H^+}=\frac{m^2_{12}}{c_\beta s_\beta}-\ums{2}v^2\la{45}$ and
$m^2_{A}=\frac{m^2_{12}}{c_\beta s_\beta}-v^2\la{5}$ are the masses squared for the charged 
Higgs and the pseudoscalar at tree-level; see e.g.\,\cite{branco.rev}.
Clearly the first and second matrices contain each a vanishing eigenvalue corresponding to the
charged ($G^\pm$) and neutral ($G^0$) Goldstone bosons in the limit where $\dm[ii]\to 0$.
We use the basis $(\phi_1^+,\phi_2^+)$, $(\sqrt{2}\im \phi_1^0,\sqrt{2}\im \phi_2^0)$
and $(\sqrt{2}\re \phi_1^0,\sqrt{2}\re\phi_2^0)$ in Eqs.\,\eqref{M:charged}, \eqref{M:odd} and
\eqref{M:even}, respectively, from the parametrization $\phi_i=(\phi_i^+,\phi_i^0)^\tp$.

Diagonalization of the tree-level part of $\cM_{\rm even}(v_i)$ defines the angle $\alpha$ 
through 
\eq{
\label{U:alpha}
U_\alpha^\dag\cM_{\rm even}^{\rm tree}U_\alpha=\diag(m^2_{H^0},m^2_{h^0})\,,
}
where
\eq{
U_\alpha=\mtrx{c_\alpha & -s_\alpha\cr s_\alpha & c_\alpha}\,.
}	
Using the same notation we can find the eigenvalues shifted by $\dm[ii]$ as
\eqali{
\label{shifted.masses}
U_{\beta_+}^\dag\cM_{\rm char}(v_i)U_{\beta_+}&=\diag(M^2_{G^+},M^2_{H^+})\,,\cr
U_{\beta_0}^\dag\cM_{\rm odd}(v_i)U_{\beta_0}&=\diag(M^2_{G^0},M^2_{A^0})\,,\cr
U_{\alpha'}^\dag\cM_{\rm even}(v_i)U_{\alpha'}&=\diag(M^2_{H^0},M^2_{h^0})\,,
}
where the angles $\beta_+,\beta_0,\alpha'$ are shifted from $\beta,\beta,\alpha$ by a small amount due to $\dm[ii]$:
\eq{
\label{shifted.angles}
\beta_+=\beta+\delta\beta_+,\quad
\beta_0=\beta+\delta\beta_0,\quad
\alpha'=\alpha+\delta\alpha.
}
The explicit forms for $\delta\beta_{+,0}$ and $\delta\alpha$ can be seen in appendix \ref{ap:dddV}.

\section{Pole masses at one-loop}
\label{sec:pole}

The previous section showed a way of ensuring that one of the minima of the effective potential
at one-loop corresponded to our vacuum with $v=246\,\unit{GeV}$ and that $\tan\beta=v_2/v_1$ could be
used as input at one-loop.
The following task to describe a realistic 2HDM at one-loop is to ensure that the SM higgs
boson mass corresponds to the experimentally measured value\,\cite{pdg}:
\eq{
\label{125}
m_{h}\big|_{\rm exp}=125.09\pm 0.24\,\unit{GeV}\,.
}
It is clear that at one-loop we cannot use this value for the tree-level parameter $m_h$ in
\eqref{input}.
Instead, we must check that the pole mass corresponds to \eqref{125}.%
\footnote{%
Another possibility is to use a more physical renormalization condition to ensure this\,%
\cite{krause}.
The renormalization of the entire theory is discussed in Ref.\,\cite{2hdm:renorm}.
}

We follow Ref.\,\cite[b]{martin} to calculate the pole masses $\hm_S$ of all the scalars $S$ including the SM higgs boson by computing the self-energies of the theory at one-loop.
We focus in this section on the CP even sector which will give rise to the pole masses of $h$ and $H$
restricted to the case $\hm_H>\hm_h$.
The self-energy for the other sectors are given in appendix \ref{ap:self-energy}.

The scalar self-couplings can be extracted from $V_0$.
Given a set of real scalars $S_i$ that interact through the quartic 
vertex $-i g_{ijkl}$ and cubic vertex $-ig_{ijk}$, the self-energy $\Pi_{ij}$ for $S_i$--$S_j$ coming from scalars in the loop with incoming momentum $p^2=s$ is given by\,\cite{martin}
\eq{
16\pi^2\Pi_{ij}^S(s)=\ums{2}\sum_{k}g_{ijkk}A(m_k)+
\ums{2}\sum_{kl}g^2_{ijkl}B(m_k,m_l,s)\,,
}
assuming we are in the basis with diagonalized masses (quadratic part of $V_0+\delta V$).
The $A$ and $B$ functions are the Passarino-Veltman functions\,\cite{PV} which, in the notation
of Ref.\,\cite{pedro}, read
\begin{eqnarray}
\label{A}
A(m) &\equiv&  m^2 \left[ \log\left(\frac{m^2}{\mu^2}\right)\,-\,1\right],\\
\label{B}
B(m_{1},m_{2},s) &\equiv&  \int^1_0 \,dt\, \log \left[ \frac{t\,m_{1}^2 \,+\, (1 - t)\,m_{2}^2 \,-\,
t(1 - t) s}{\mu^2}\right].
\end{eqnarray}
The $A$ function represents the one-loop graph with one quartic vertex (tadpole) and the $B$
function represents the one-loop graph with two vertices and two internal lines.
Hence the factors $1/2$ are the symmetry factors that appear in front of the Feynman diagrams for identical fields.
These functions appear after renormalizing these diagrams using the $\msb$ prescription.
Simplifications for vanishing $s$ can be found in Ref.\,\cite{pedro}.
The contributions coming from gauge bosons and fermions in the loop are also shown in appendix
\ref{ap:self-energy} and all the necessary cubic and quartic couplings of the theory are explicitly shown in appendix \ref{ap:cubics}.

For example, the SM higgs self-energy is given by
\eqali{
16\pi^2\Pi^S_{hh}(s)&=
\ums{2}\sum_{S=G^0,A,H,h}g_{h^2S^2}A(M_S)
+\sum_{S=G^+,H^+}g_{h^2S\bar{S}}A(M_S)
\cr
&\quad
+\ums{2}\sum_{S,S'=G^0,A,H,h}g_{hSS'}^2 B(M_S,M_{S'},s)
+\sum_{S,S'=G^+,H^+}|g_{hS\bar{S'}}|^2 B(M_S,M_{S'},s)\,,
}
where $g_{h^2h^2}=g_{h^4}$ and $g_{hh^2}=g_{h^3}$ are the quartic and cubic self-couplings for $h$
and all $M_S$ refer to $M_S(v_i)$.
However, the mixing to the other CP even scalar $H$ in $\Pi_{Hh}$ cannot be neglected.

Due to charge and CP conservation, the self-energy for the different scalars decouple
into separate pieces for the CP even, CP odd and charged sectors.
The self-energy for the CP even sector is given by the matrix
\eq{
\Pi^{\rm even}(s)=\mtrx{\Pi_{HH}(s) & \Pi_{Hh}(s) \cr \Pi_{hH}(s) & \Pi_{hh}(s)}\,.
}
The one-loop pole squared masses $\hm_H^2,\hm_h^2$ for the CP even scalars $H,h$ are the solutions $s_k$
for\,\cite{martin}
\eq{
\det\mtrx{
	M_H^2 +\re\Pi_{HH}(s_k) -s_k & \re\Pi_{Hh}(s_k)\cr
	\re\Pi_{hH}(s_k) & M_h^2+\re\Pi_{hh}(s_k)-s_k }
=0\,,
}
where we take only the real part of the self-energy because we are not interested in
the decay widths.
The $\hm_h^2$ corresponds to the solution that continually approaches $M_h^2$ in the limit 
where we turn off the interactions.
A similar consideration applies to all pole squared masses.

\section{Numerical survey}
\label{sec:numerical}

We describe here the procedure we used to survey the models. For definiteness, we
adopted the relevant fixed parameters of the SM to be
\eqali{
g = 0.6483; \quad g'=0.3587;
\quad v = 246.954\;\unit{GeV};
\quad y_{t} = 0.93697;
\quad y_b=0.023937\,. 
}
The first four values were taken from Refs.\,\cite{strumia,martin.14} as the running parameters
of the SM 
at the top pole mass $\mu=173.34\,\unit{GeV}$ and the bottom Yukawa was adapted from its mass value
$m_b=4.18\,\unit{GeV}$\,\cite{pdg}. Among these parameters, only the top Yukawa
appreciably affects the one-loop effective potential together with the quartic
scalar self-couplings. We use a fixed renormalization scale
$\mu=300\,\unit{GeV}$ for all calculations and note that the running of $y_t$ from
the top mass scale only amounts to a small difference. The running for the rest of
the parameters are even less relevant.
We remark that any choice of the renormalization scale is allowed, since the difference in depth of the potential at two extrema is a renormalization scale-independent quantity\,\cite{scale}. However, from a practical point of view, it is desirable that the logarithms in the effective potential do not become too ``large'' so as to lead to numerical instabilities\,\cite{casas}. We have checked our calculation for some different values of the renormalization scale and the value of $\mu=300\,\unit{GeV}$ proved to be a stable choice.

Among the input parameters in \eqref{input}, we fixed the standard vev $v$ as above and
took the rest of the parameters randomly in the range shown in the first row of
Table\;\ref{table:ranges},
restricted to $m_H>m_h$.
After checking for simple perturbativity and bounded from below conditions at tree-level,%
\footnote{%
Since we work with a fixed renormalization scale where the one-loop corrections are not large,
we expect the tree level relations to be valid to a good 
approximation\,\cite{lambda}.
For bounded from below conditions, this is a conservative choice as one-loop 
corrections may enlarge the possible parameter space\,\cite{staub}.
}
we picked only the points where the shifted masses squared $M_S^2(v_i)$ were positive
and the solutions for the shift $\dm[ii]$ in \eqref{eq:deltam} were real.
Then we further selected only the points where the pole mass $\hat{m}_h$ calculated as in
Sec.\,\ref{sec:pole} fell in the experimental range of \eqref{125}.%
\footnote{%
The exact adopted procedure differs slightly with respect to the range of
$m_h$: instead of post-selecting only the values of $m_h$ for which the pole mass
coincided with the experimental range, we randomly selected the input $m_h$ in the
range $[0,200]$\,\unit{GeV} and then later varied only this value searching for the correct
pole mass $\hat{m}_h$. If a solution were found, we kept that point.
This procedure resulted in the approximately homogeneous distribution of $m_h$
in the range shown in Table\,\ref{table:ranges}.
This modification speeded up the generation of points.
}
In this way, we generated the sample G with 294437 points among which 4525 had two minima
at one-loop, 17 of which were non-regular.
To find the second minimum, we explicitly minimized the real part of the effective
potential \eqref{V0+V1} starting from the non-standard minimum at tree-level
and then retained only the points where the value of the potential at that minimum was
real\,\cite{weinberg.wu}.
Given the small number of non-regular points,
we generated another sample denoted as NR focusing only on non-regular points by
imposing large $t_\beta$ as in the second row of Table\;\ref{table:ranges}; other ranges
were kept the same, except for $m^2_{12}$ which was chosen positive.
Sample NR thus contained 185905 points among which 1563 had two minima at one-loop.
Hereafter, unless explicitly specified, we will consider only the joint sample of G and NR. However, we would like to emphasize that, even though samples G and NR have a similar number of points, the parameter space scanned by sample NR represents only a small portion of the parameter space probed by sample G. This justifies our definition of non-regular points.
\begin{table}[h]
\eq{\nonumber
\begin{array}{|c|c|c|c|c|c|c|c|}
\hline
\text{Sample} &
t_\beta & \cos(\beta-\alpha) & m_{H^+}\, (\unit{GeV})& m_{A}\,(\unit{GeV})&
m_{H}\,(\unit{GeV})& m_{h}\,(\unit{GeV}) & m^2_{12}\,(\unit{GeV^2})
\cr
\hline
\text{G} &
\phantom{2}1 \div 50 & -0.1 \div 0.1 & 90 \div 1000& 90\div 1000& 90\div 1000&
0\div 150 &
-6000 \div 6000
\cr
\hline
\text{NR} &
20 \div 50 & \text{--} & \text{--} & \text{--} & \text{--} & \text{--} &
\phantom{-600}0 \div 6000
\cr
\hline
\end{array}
}
\caption{Input parameter ranges for the general sample (G) and non-regular points (NR).
The dash denotes the same range as in the previous row.
}
\label{table:ranges}
\end{table}

After the selections described above, the input parameters $m_{H^+},m_A,m_H$ get 
roughly confined to the range $[90,500]$ and
the non-standard higgsses acquire 
pole masses in a similar range with slightly smaller maximal value for
$\hat{m}_{H^+}$.
In contrast, the distribution for $t_\beta$ is homogeneous in the range of Table\,\ref{table:ranges}
for the whole sample but, as we select only the points with two minima at one-loop, 
it gets separated into two ranges,
$1\le t_\beta \lesssim 3.8$ for the regular minima, and 
$24\lesssim t_\beta \le 50$ for the non-regular ones.

To check our numerically implemented formulas, we performed the following consistency checks:
\begin{enumerate}
\item Vanishing of the pole masses for the Goldstone bosons $G^{\pm},G^0$ for zero
external momentum for the three cases where we successively add the one-loop scalar, vector boson and fermion contributions.
\item Equality of the pole masses for the CP even higgsses $H,h$ for zero external momentum
and the eigenvalues of the explicitly calculated second derivative matrix of the effective potential in the real neutral directions \eqref{varphii}
in all three cases of successive addition of the one-loop scalar, vector boson and fermion
contributions.
\end{enumerate}

\section{Results}
\label{sec:results}

\begin{figure}[h]
\includegraphics[scale=0.4]{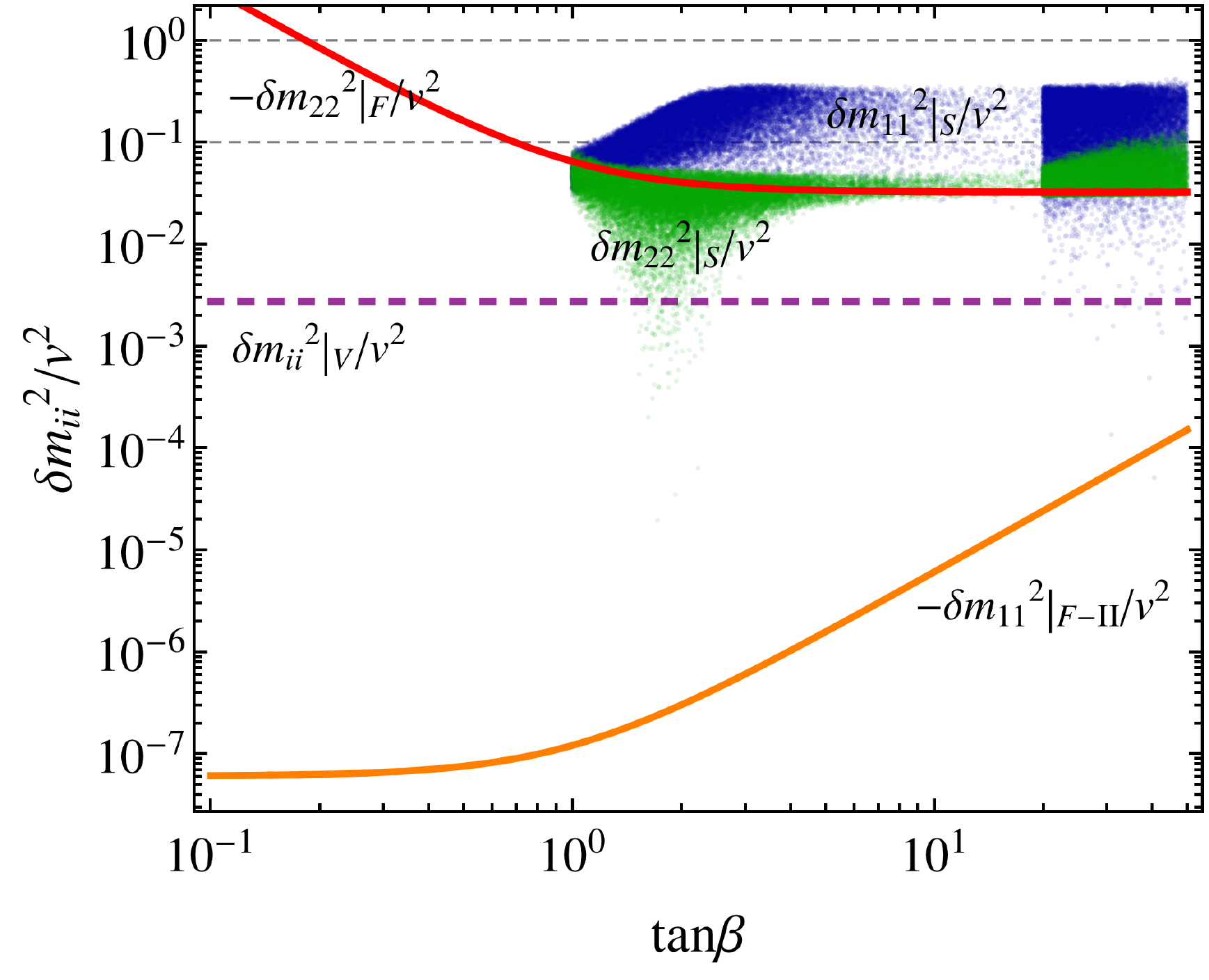}
\caption{
Different contributions for $\dm[ii]$ in \eqref{eq:deltam}.
The scalar contributions are positive for $\dm[11]$ (blue points) and mostly 
positive for $\dm[22]$ (green points) while the fermions contribute negatively.
The fermion contribution to $\dm[22]$ (red curve) is practically the same for the
type I and II models while the contribution to $\dm[11]$ (orange curve) applies only
to the type II case but vanishes for the type I case.
The contributions from the gauge bosons are shown in the purple dashed line.
}
\label{fig.3}
\end{figure}

\begin{figure}[h]
\includegraphics[scale=0.4]{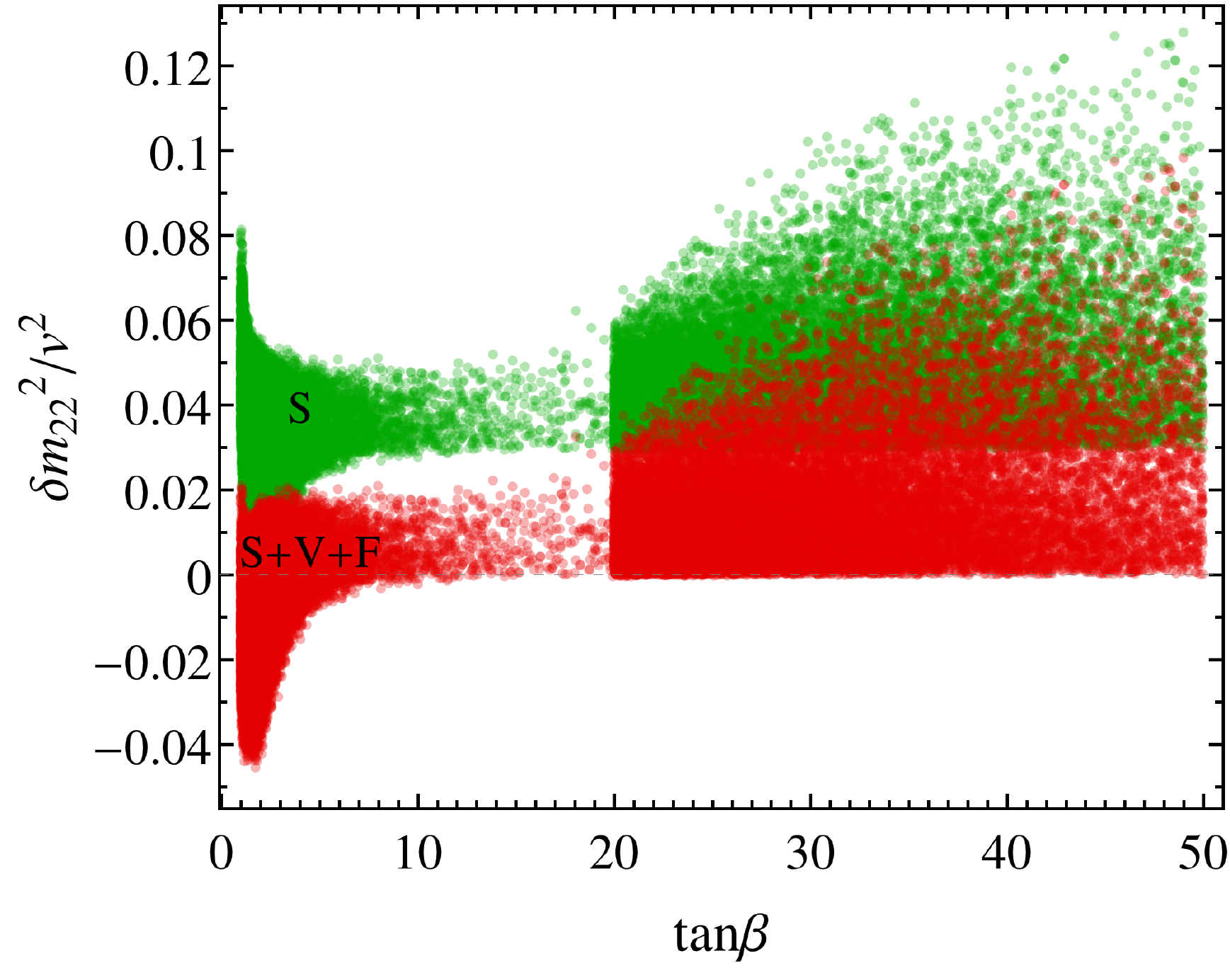}
\caption{
Different contributions to $\dm[22]$ in \eqref{eq:deltam}: the green points quantify 
the scalar contribution only whereas the red points consider all the contribution in the type I or 
II model.
}
\label{fig.4}
\end{figure}

Let us first quantify the shifts $\dm[ii]$ in \eqref{eq:deltam} for the different contributions.
The contribution coming from the gauge bosons only depends on the value of $v$ and,
for fermions, it depends on $v$ and $\beta$.
The scalar contribution depends on many parameters coming from the scalar potential.
The dependence of the different contributions on $\tan\beta$ are shown in Fig.\,\ref{fig.3}.
We can see that the dominant contribution for $\dm[11]$ comes from the scalars whereas
$\dm[22]$ also has large positive contributions from scalars but they are partly canceled
by the negative contribution from fermions (top).
Such a partial cancellation in $\dm[22]$ can be clearly seen in Fig.\,\ref{fig.4} where we show
the contribution only from scalars (green points) and all the contributions (red points).
The orange curve in Fig.\,\ref{fig.3}, which quantifies the bottom contribution to
$\dm[11]$ in the type II model, shows that it is negligible in the range of $\tan\beta$
we are interested in and our calculation that uses the type I model applies equally well 
to the type II case.
All the different contributions are calculated using Eq.\,\eqref{extremum:1-loop}
and the scalar contribution in particular depends on the $\dm[ii]$ themselves and these
are taken as the total contributions.
The remaining contributions of gauge bosons (purple dashed curve in Fig.\,\ref{fig.3}) are much smaller. 
The contribution from scalars are calculated using the whole sample (G\,$+$\,NR) described in Sec.\,\ref{sec:numerical} which also includes the points with only one vacuum.

\begin{figure}[h]
\includegraphics[scale=0.38]{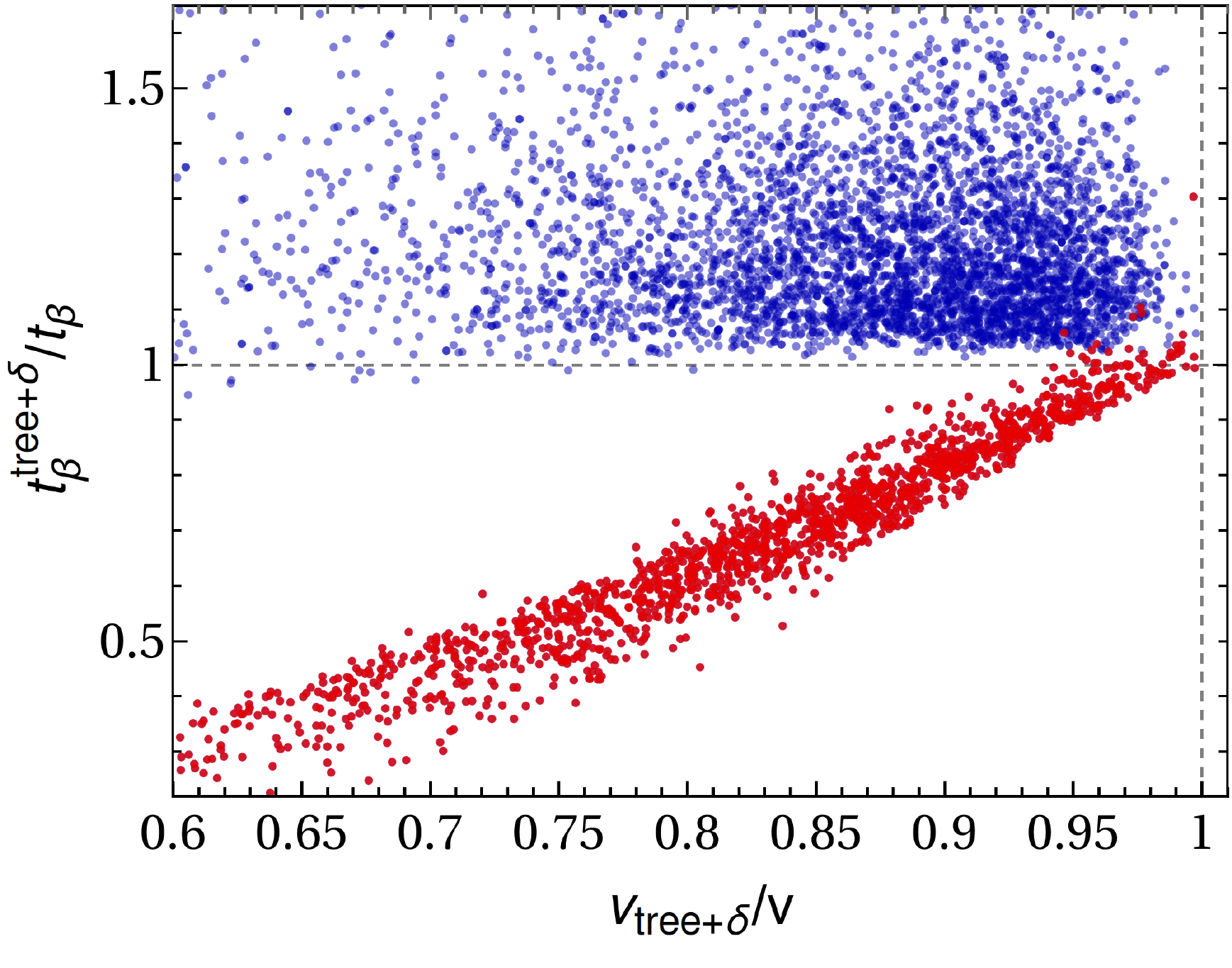}
\caption{Variation of $t_\beta$ and $v$ when shifting the quadratic parameters $m_{ii}^{2}$ of
\eqref{pot:tree} by $\dm[ii]$. The blue (red) points represent (non-)regular points. 
}
\label{fig.5}
\end{figure}

The deviation of the location of our vacuum when we add $\delta V$ do $V_0$ is illustrated
in Fig.\,\ref{fig.5} where we show the ratio of $t_\beta$ for $V_0+\delta V$
($t_\beta^{\rm tree+\delta}$) to that of $V_0$ ($t_\beta$) against the ratio of the vev for $V_0+\delta V$ ($v_{\rm tree+\delta}$) to that of $V_0$ ($v$).
We only show the points with two coexisting minima and divide the points between the regular ones (blue) and the non-regular ones (red).
We can see that as $m_{ii}^{2}$ get shifted by $\dm[ii]$ all points with two minima have their vevs decreased while $t_\beta$ mostly increases for the regular points and mostly decreases for the non-regular points.
Note that the non-regular points only consider large $t_\beta$ whereas the regular
points only include moderate $t_\beta$ roughly up to $3.8$.
As the location of $(v_1,v_2)$ for our vacuum is the same for $V_0$ and the one-loop
corrected potential \eqref{V0+dV+V1l} we can also interpret this plot as the modification
of the vev location of $V_0+\delta V$ compared to $V_0+\delta V+V_{1l}$.
If we had considered all the points including the points with only our vacuum,
the majority of points would follow the behavior of the regular points in blue.

\begin{figure}[h]
\includegraphics[scale=0.4]{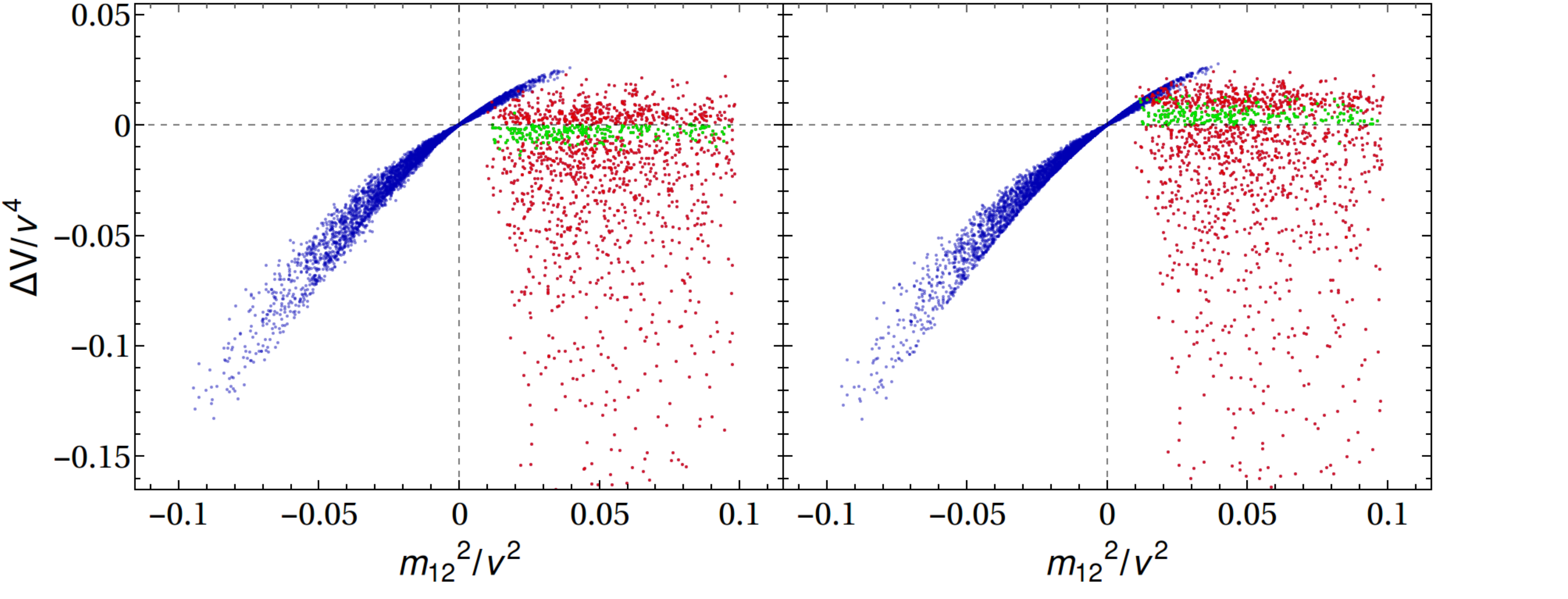}
\caption{Potential difference of the two coexisting minima against $m_{12}^{2}$. 
\textbf{Left:} Only the tree level potential (\ref{pot:tree}) is considered. This 
plot should be compared with Fig.\;\ref{fig.1}. \textbf{Right:} The full 1-loop 
effective potential (\ref{V0+dV+V1l}) is considered. The green points 
represent a change in sign of the tree level prediction for the potential 
difference.
}
\label{fig.6}
\end{figure}

For the case of coexisting vacua, we can see a clear difference in behavior between 
the regular points and the non-regular ones in Fig.\,\ref{fig.6} where we plot
the potential difference with respect to $m^2_{12}$.
The left panel shows these quantities using the tree-level potential $V_0$
while the right panel is the same plot for the one-loop potential \eqref{V0+dV+V1l}.
We can clearly see that the potential difference goes continuously to zero as $m^2_{12}\to 0$
for the blue points representing the regular minima.
Moreover, for positive $m^2_{12}$ our vacuum is guaranteed to be the global minimum
in both tree level or one-loop potential.
This behavior is opposite for negative $m^2_{12}$.
The reason is that only $m^2_{12}$ breaks explicitly the $\ZZ_2$ symmetry of the theory and it
controls the degeneracy breaking of the spontaneously breaking minima.
This behavior is not followed by the non-regular minima that do not have degenerate
minima in the $m^2_{12}\to 0$  limit.
Some points (green) in which our vacuum is not the global one at tree-level even get inverted
and become the global minimum as the one-loop corrections are added.
The same conclusion is reached if we had compared the one-loop potential to $V_0+\delta V$
instead: some cases where our minimum is not the deepest at tree-level becomes the global minimum
at one-loop.

\begin{figure}[h]
\includegraphics[scale=0.42]{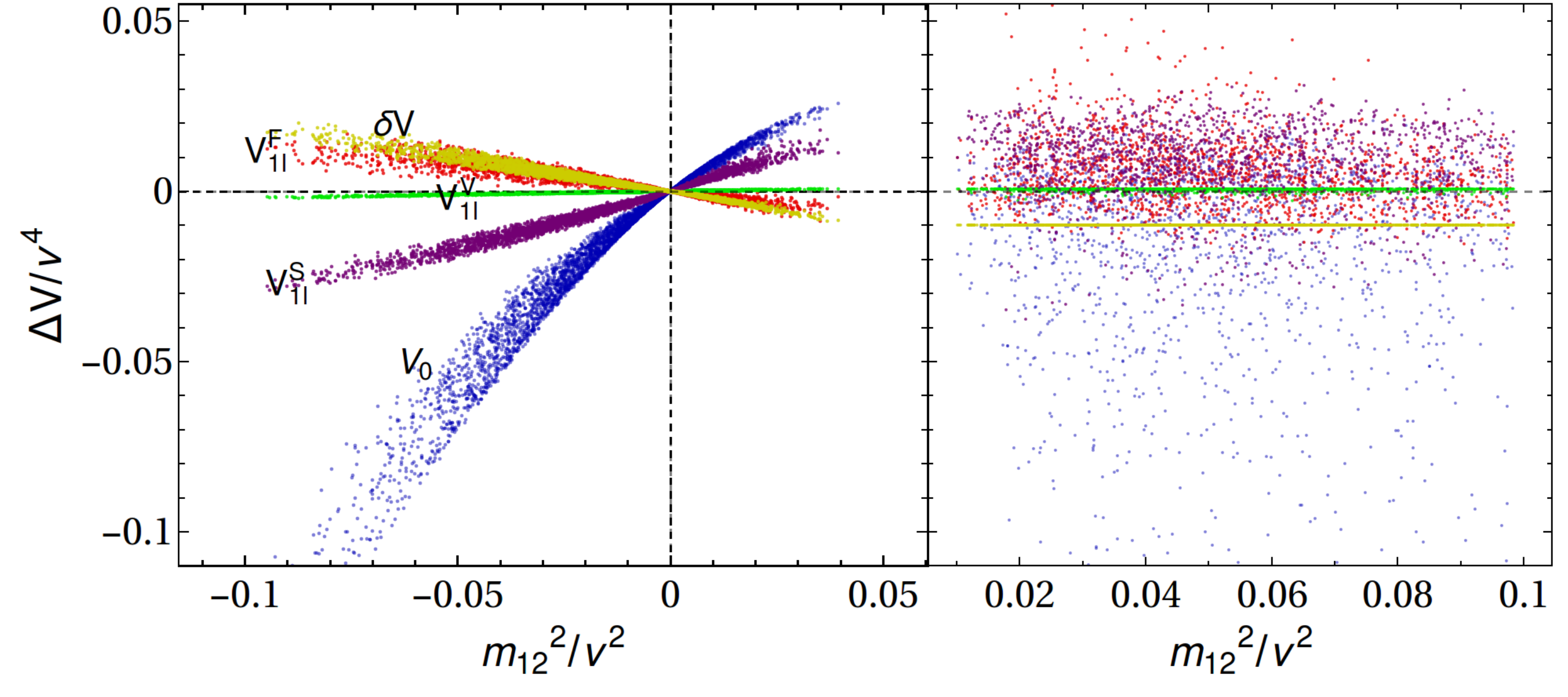}
\caption{
Different contributions coming from the tree level potential (blue), $\delta V$ (yellow), fermions (red), scalars (purple), and vectors bosons (green) to the one-loop potential difference of the right panel of Fig.\,\ref{fig.6}.
In the left plot only regular points are considered while the right plot shows the behavior of the non-regular ones.
}
\label{fig.7}
\end{figure}

\begin{figure}[h]
\includegraphics[scale=0.45]{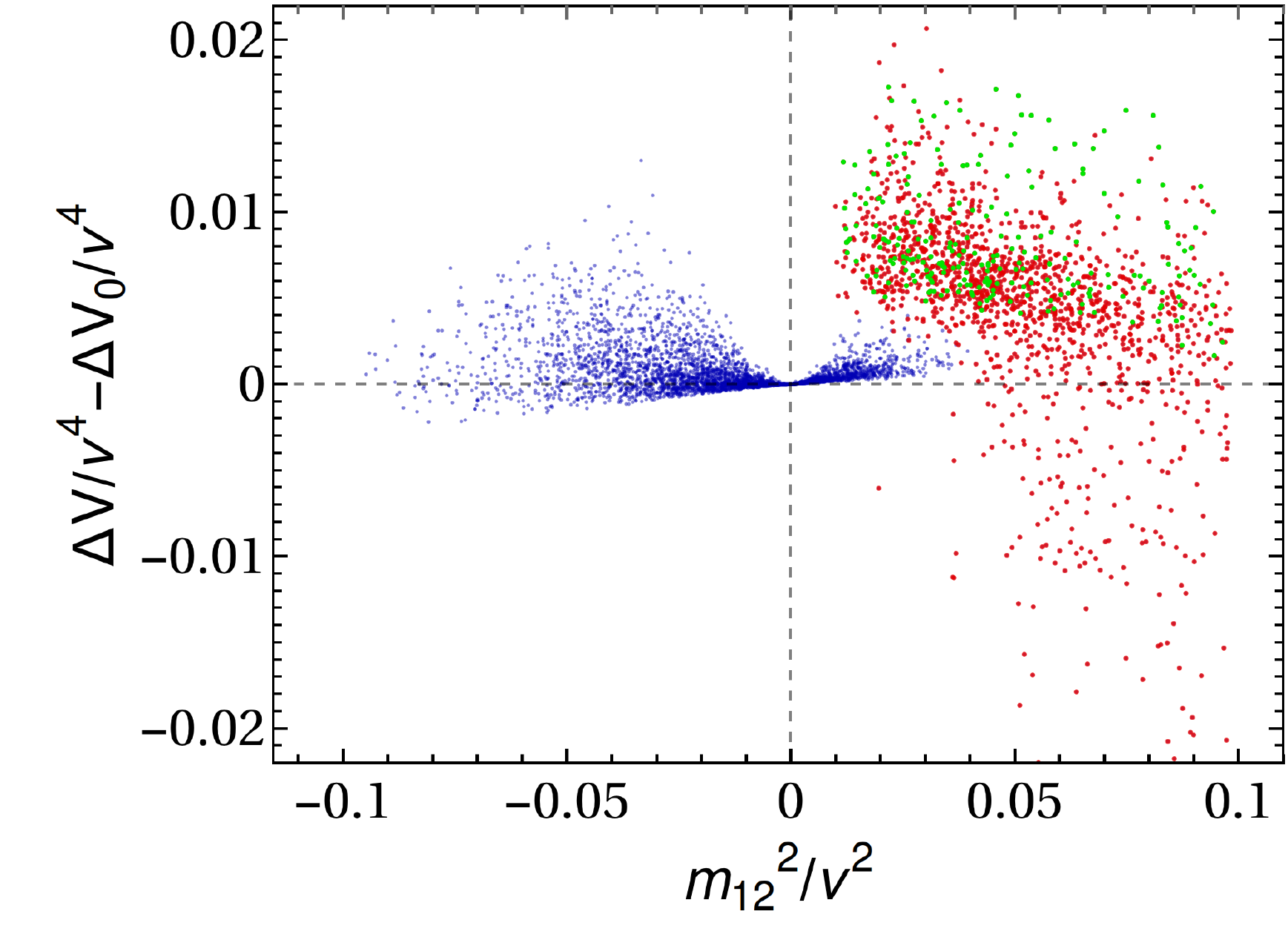}
\caption{
One-loop correction to the potential difference relative to the tree level potential difference.
The color coding follows Fig.\,\ref{fig.6}.
}
\label{fig.8}
\end{figure}

We can have an idea of the different one-loop contributions in Fig.\,\ref{fig.7}
where we separate the potential difference in Fig.\,\ref{fig.6} into its different
contributions.
Regarding the regular points (left plot), it can be seen that the one-loop potential difference
is almost entirely due to the tree-level contribution (blue) since
the contributions from $\delta V$ (yellow), fermions (red) and scalars (purple)
approximately cancel each other while the contribution from gauge bosons (green)
is negligible compared to the others.
This behavior justifies our choice for the renormalization scale.
For the non-regular points (right plot), no clear pattern emerges.
We can also see in Fig.\,\ref{fig.8} that there are points where the potential difference is raised as well as points where it is lowered by the one-loop corrections
for both regular and non-regular points.

Considering that the discriminant \eqref{discriminant} test is applicable
for all cases where at least one normal vacuum is known, we can investigate if it
is still a good predictor for the one-loop potential with two coexisting minima.
We can adapt the tree-level discriminant to one-loop in the following three ways:
\eqali{
\label{D:123}
D_1&=\text{Eq.\,}\eqref{discriminant}\,,
\cr
D_2&=\text{Eq.\,}\eqref{discriminant}\big|_{m^2_{ii}\to m^2_{ii}+\dm[ii],\,v\to v^{(0)},
\,\beta\to \beta^{(0)}}\,,
\cr
D_3&=\text{Eq.\,}\eqref{discriminant}\big|_{m^2_{ii}\to m^2_{ii}+\dm[ii]}\,.
}
The quantity $D_1$ is the discriminant calculated with $V_0$ as the whole potential
while $D_2$ is calculated by using $V_0+\delta V$ in \eqref{V0+dV+V1l}.
In the latter case, the quadratic parameters $m^2_{ii}$ get shifted and the location
of the minimum, denoted above as $(v^{(0)},\beta^{(0)})$ or as 
$\rm tree+\delta$ (superscript/subscript) in Fig.\,\ref{fig.5},
do not coincide with the ones used as input, denoted as $(v,\beta)$.
The last adaptation $D_3$ considers the shifts in the quadratic parameters but keeps the vevs as
$(v,\beta)$.
We will test here if any of these discriminants are capable of distinguishing if our vacuum
is the global one at one-loop only using parameters at tree-level.%
\footnote{%
Strictly speaking, some calculation at one-loop is required for some of these quantities
depending on how the calculation is set up.
Using the splitting of the potential in the form \eqref{V0+dV+V1l}, $D_1$ is the most
natural quantity to use and there is no one-loop calculation required.
If $V_0+\delta V$ is considered as the potential at tree-level,
$D_2$ or $D_3$ are the natural quantities depending on which minimum is taken.
}

\begin{figure}[h]
\includegraphics[scale=0.37]{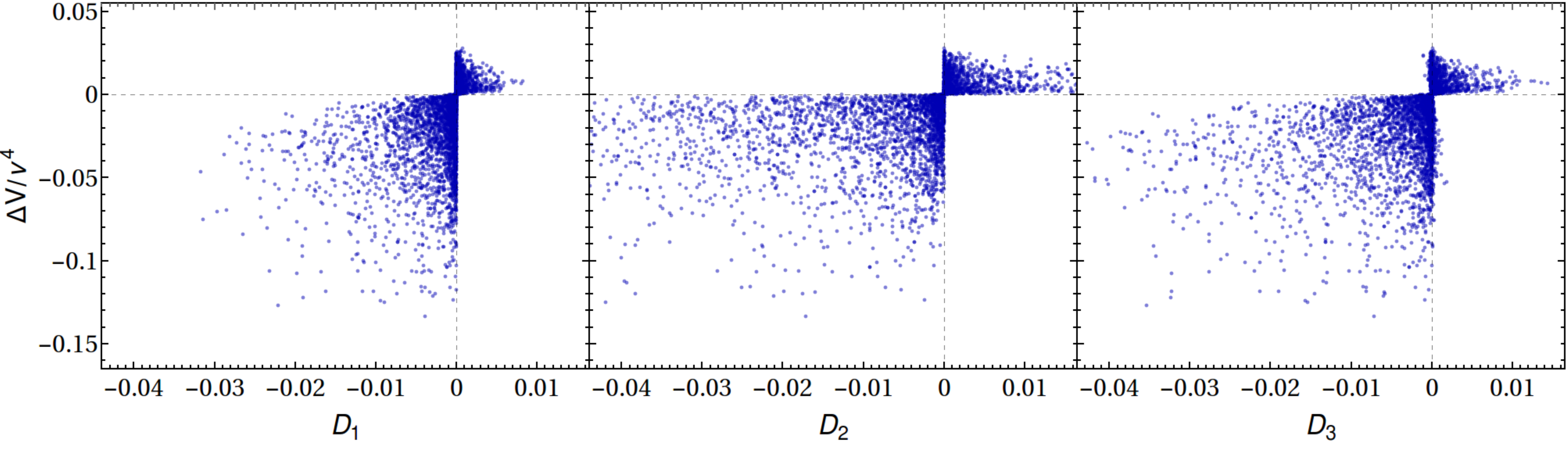}
\caption{Comparison of the three discriminants proposed in \eqref{D:123} in the case of 
regular points. 
}
\label{fig:D_regular}
\end{figure}

For the regular coexisting minima, Fig.\,\ref{fig.6} shows that
$m^2_{12}$ is already a good predictor of the global minimum, but we can test the
discriminants in \eqref{D:123}.
The result of this test is shown in Fig.\,\ref{fig:D_regular} where the potential
difference at one-loop is plotted against the three discriminants.
We can see in the left and middle plots that $D_1$ and $D_2$ correctly predict 
the global minimum of the one-loop effective potential while the right plot shows that 
$D_3$ fails for more than 10$\%$ of the points.

\begin{figure}[h]
\includegraphics[scale=0.37]{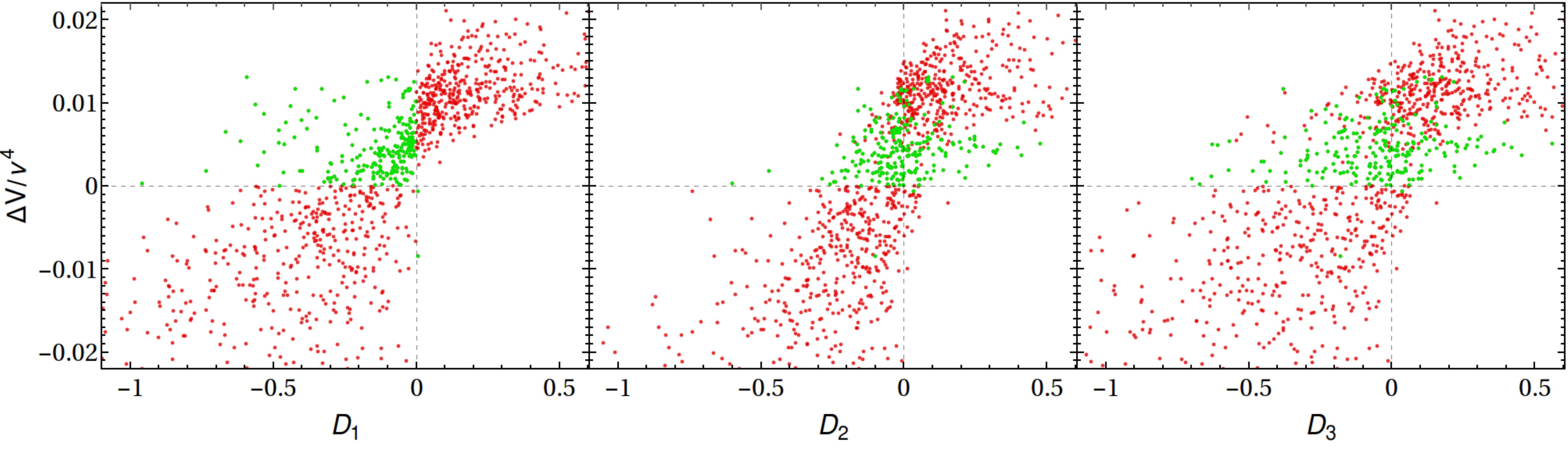}
\caption{Comparison of the three discriminants proposed in \eqref{D:123} in the case of 
non-regular points.
The green points mark the cases where $D_1$ predicts the opposite depth difference.
}
\label{fig:D_nonregular}
\end{figure}

In constrast, the failure of \textit{all} the discriminants \eqref{D:123} for the non-regular points (of samples G and NR) can be seen in
Fig.\,\ref{fig:D_nonregular} which shows the potential depth difference as a function of the discriminants, similarly to the regular points in Fig.\,\ref{fig:D_regular}; note that the
horizontal scale is very different.
The green points in the second and fourth quadrants mark the cases where the discriminant $D_1$ predicts the opposite behavior at one-loop.
We can clearly see that all discriminants fail for a significant portion of points,
not necessarily for the same ones.
From the property of $D_1$, we could have seen its wrong prediction in Fig.\,\ref{fig.6}
as well.

At last, to make sure that the points for which the discriminant test fails include
phenomenologically realistic points, 
we have checked the viability of the red/green points in Fig.\,\ref{fig.6}
by considering 
phenomenological constraints implemented in the 2HDMC code\,\cite{2HDMC}. We found 
that in the case of the type I model around half of the green points are allowed by 
experimental constraints such as the S, T, U precision electroweak parameters and data from 
colliders implemented in the HiggsBounds and HiggsSignals packages\,\cite{HB,HS}, 
while in the type II model they are all excluded.
For the type II model we have also included constraints coming from $R_{b}$ measurements\,\cite{constrain_R:H+} as well as $B$ meson decays\,\cite{constrain:H+} which prove to be very strong by setting $m_{H^+}> 480\,\unit{GeV}$ independently of the value of $\tan\beta$. For type I, since the red/green points have $\tan \beta>10$, these constraints do not impose any further restrictions\,\cite{constrain_g:H+}.
We also checked that all points still respect simple bounded from below and perturbativity 
constraints\,\cite{branco.rev}.

\section{Conclusions}
\label{sec:conclusions}

We have studied the one-loop properties of the real Two-Higgs-doublet model with softly broken $\ZZ_2$ with respect to the possibility of two coexisting normal vacua.
The softly broken nature must remain at one-loop and the case of two coexisting normal minima
can be classified into two very distinct types depending on their nature in the vanishing
$m^2_{12}$ limit:
regular minima that spontaneously break the symmetry and the non-regular minima (or minimum)
that preserve the symmetry, i.e., they are inert-like in the symmetric limit.
Since in the first case the two minima are degenerate in the symmetry limit, even at one-loop,
they are connected by the $\ZZ_2$ symmetry and then they should differ only by the sign of $v_2$.
After the inclusion of the $m^2_{12}$ term, the sign of $v_2$ continue to be opposite
for our vacuum and the non-standard one so that the two regular minima
are found in the first and fourth quadrants in the $(v_1,v_2)$ plane.
In contrast, the non-regular minima deviate from the inert-like minima and both deviate
to the first quadrant when $m^2_{12}$ is positive.

The vacua that spontaneously break $\ZZ_2$ in the $m^2_{12}\to 0$
limit behave rather regularly and we can distinguish which coexisting minimum is the
global one by just examining the sign of $m^2_{12}$: when it is negative
our vacuum is only a metastable local one and the opposite is true if it is positive.
For this type of coexisting vacua, the discriminant at tree level [$D_{1,2}$ in
Eq.\,\eqref{D:123}] is still a good predictor of the nature of the minimum it is calculated
with.

For the non-regular coexisting vacua, $m^2_{12}$ is positive in the convention that
our vacuum has both vevs positive and it cannot be used as an indicator.
At tree level, the discriminant of Ref.\,\cite{panic.vacuum} is a very convenient way 
of testing if our vacuum is the global one because only the location of our minimum
is required.
However, at one-loop,
this discriminant is not a precise indicator of which minima is the global one.
We have found realistic cases where our vacuum is not the global minimum at tree-level but it becomes
the global one after the addition of the one-loop corrections.
Few cases for the opposite behavior were also found.
As the discriminant effectively distinguishes the sign of the potential difference between
the coexisting minima at tree-level, the latter itself is also a not good indicator for the non-regular minima.
This is reminiscent of the exact $\ZZ_2$ symmetric case (inert model) investigated in
Ref.\,\cite{pedro}.
On the other hand, we were unable to find a discriminant that works for both regular and non-regular 
minima at one-loop. Finding a simple and precise criterion for global minimum does not seem to be an easy task as that was not achieved even in the simpler exact $\ZZ_2$ limit.
We also emphasize that for our parametrization which enforces our vacuum to be a minimum
from the start, and for the chosen generic ranges of parameters (sample G),
the occurrence of non-regular minima, as suggested by its name, is much rarer:
only $38\%$ correspond to the non-regular cases and, among then, only $0.3\%$
are coexisting minima.

In summary, the soft-breaking term $m^2_{12}$ controls the lifting of the degeneracy of the regular coexisting minima
(moderate $\tan\beta$)
even at one-loop and can be used as the sole indicator of which minimum is the global one.
That is not true when two coexisting non-regular vacua
(large or small $\tan\beta$) exist: the discriminant that
is a precise indicator at tree-level is not reliable at one-loop and explicit calculation of the potential depths must be carried out.

\acknowledgements

A.L.C. acknowledges financial support by Brazilian CAPES (Coordena\c{c}\~{a}o de Aperfei\c{c}oamento de Pessoal de N\'{i}vel Superior)
and 
C.C.N acknowledges partial support by Brazilian Fapesp grant 2014/19164-6 and
2013/22079-8, and CNPq 308578/2016-3.
The authors thank Pedro Ferreira for useful discussions while this work was developed.

\appendix
\section{Deviation of a potential minimum}
\label{ap:deviation}

Take a potential $V_0(\varphi)$ depending on real scalar fields $\varphi_i$ for 
which we know a minimum (extremum) $\varphi_i=\bvphi_i$ satisfying
\eq{
\label{min:V0}
\frac{\partial V_0(\bvphi)}{\partial \varphi_i}=0\,.
}
We want to quantify how the location of the minimum and the value of the 
potential deviate when we add a small perturbation $U$ on the potential as
\eq{
V=V_0+U\,.
}
We assume $V_0$ has no flat direction around $\bvphi$.

We first quantify the deviation of the location of the minimum as
\eq{
\varphi_i=\bvphi_i+\delta\varphi_i\,.
}
The derivative of the perturbed potential gives
\eq{
\frac{\partial V_0(\bvphi+\delta\varphi)}{\partial \varphi_i}=
\frac{\partial V_0(\bvphi)}{\partial \varphi_i}+
\frac{\partial U(\bvphi)}{\partial \varphi_i}+
\frac{\partial^2 V_0(\bvphi)}{\partial \varphi_i\partial \varphi_j}\delta \varphi_j
+O(\delta^2)\,.
}
Since the first term vanishes due to \eqref{min:V0}, we get the deviation to first 
order
\eq{
\label{delta.phi}
\delta\varphi_j= -\cM^{-1}_{ji}U_i\,,
}
where $U_i\equiv\partial U(\bvphi)/\partial\varphi_i$ and
$\cM_{ij}=\frac{\partial^2 V_0(\bvphi)}{\partial\varphi_i\partial\varphi_j}$
is the squared-mass matrix around $\bvphi$.

The deviation in the value of the minimum can be equally expanded around $\bvphi$ as
\eq{
V(\bvphi+\delta\varphi)=V_0(\bvphi)+U(\bvphi)-\ums{2}U_i\cM^{-1}_{ij}
U_j+O(\delta^3)\,,
}
after using \eqref{delta.phi}.
Generically the third term contributes to deepen the potential.
That is the dominant contribution when the perturbation vanishes on the unperturbed 
minimum: $U(\bvphi)=0$. The latter happens for the $m^2_{12}$ term when perturbing the inert-like minima while for the spontaneously broken $\ZZ_2$ minima the dominant term is the 
second, linear in the perturbation.

\section{Finding more than one normal vacua}
\label{ap:NV'}

Let us rewrite the minimization equations in Eq.\,\eqref{extremum:tree} as
\eq{
\label{mineq:A}
A_\zeta X=M\,,
}
where 
\eq{
A_\zeta=\mtrx{\la1 & \la{345}-\zeta\cr \la{345}-\zeta & \la2 }\,,\quad
X=\mtrx{v_1^2\cr v_2^2}\,,\quad
M=-2\mtrx{m^2_{11}\cr m^2_{22}}\,,
}
and $\zeta\equiv 2m^2_{12}/v_1v_2$ depends on the vevs.

To find solutions $(v_1,v_2)$ of \eqref{mineq:A} for $m^2_{12}\neq 0$, we first need 
an equation for $\zeta$.
For that end, we can formally write \eqref{mineq:A} as
$X=A_\zeta^{-1}M$ and equate $v_1^2v_2^2=4m^4_{12}/\zeta^2$.
We obtain
\eq{
\frac{m^4_{12}}{\zeta^2}\big[(\la{345}-\zeta)^2-\la1 \la2\big]^2
=[m^2_{11}\zeta-\mu_1][m^2_{22}\zeta-\mu_2]\,,
}
where $\mu_1=\la{345}m^2_{11}-\la1 m^2_{22}$ and $\mu_2=\la{345}m^2_{22}-\la2 
m^2_{11}$.
The $m^2_{12}\to 0$ limit is taken as $\zeta\to 0$ and $\frac{m^4_{12}}{\zeta^2}\to 
v_1^2v_2^2/4$.
When $m^2_{12}\neq 0$, we can find the possible values for $\zeta$ (extrema) from 
the quartic equation
\eq{
\label{sol.1}
m^4_{12}\big[(\la{345}-\zeta)^2-\la1 \la2\big]^2
=\zeta^2[m^2_{11}\zeta-\mu_1][m^2_{22}\zeta-\mu_2]\,.
}
We know there are at most four real solutions coinciding with the maximal number of 
extrema in this case\,\cite{ivanov:mink}.
We also can see that the root $\la{345}+\sqrt{\la1\la2}$ from the lefthandside is always 
positive due to bounded below conditions.
Possible solutions for \eqref{mineq:A} depends on the sign of
$\zeta$ and we allow both signs for $v_1v_2$.
We can see that for the same $\zeta$,
flipping the sign of $m^2_{12}$ is equivalent to flipping the sign of $v_1 v_2$.

Once some solution $\zeta=\zeta_0$ is found, we can find the vevs from the relation 
$X=A_\zeta^{-1}M$ or, explicitly,
\eqali{
v_1^2&=
\frac{2(m^2_{22}\zeta-\mu_2)}{(\la{345}-\zeta)^2-\la1\la2}
~>0\,,
\cr
v_2^2&=
\frac{2(m^2_{11}\zeta-\mu_1)}{(\la{345}-\zeta)^2-\la1\la2}
~>0
\,.
}
After ensuring these expression are positive, we can extract $v_1,v_2$ and 
$\tan\beta$ with the sign convention where $v_1>0$ and
\eq{
\sign(v_2)=\sign(\zeta m^2_{12})\,.
}

\section{Matrix of second derivatives}
\label{ap:ddV}

The mass matrices at tree-level for the charged, CP-odd and CP-even scalar sectors prior to 
imposing the minimization conditions are given respectively by
\eq{
\cM_{\rm char}(\varphi_i)=
\inner{\frac{\partial^2V_0}{\partial\phi_a^{-}\partial\phi_b^+}}=
\left(
\begin{array}{cc}
 \ums{2}\la1 \varphi_1^2+\ums{2}\lambda_3\varphi_2^2 & \frac{1}{2}\la{45}\varphi_1\varphi_2 \\
 \frac{1}{2}\la{45}\varphi_1\varphi_2 & \ums{2}\la{2}\varphi_2^2+\ums{2}\la3 \varphi_1^2 \\
\end{array}
\right)        
+(\text{quadratic})\,,        
}
\eq{
\cM_{\rm odd}(\varphi_i)=
\inner{\frac{\partial^2V_0}{\partial\eta_a\partial\eta_b}}
=
\left(
\begin{array}{cc}
 \ums{2}\la1 \varphi_1^2+\ums{2}\bla{345}\varphi_2^2 & \la{5}\varphi_1\varphi_2 \\
 \la{5}\varphi_1\varphi_2 & \ums{2}\la{2}\varphi_2^2+\ums{2}\bla{345} \varphi_1^2 \\
\end{array}
\right)                   
+(\text{quadratic})\,,
}
\eq{
\cM_{\rm even}(\varphi_i)=
\inner{\frac{\partial^2V_0}{\partial\rho_a\partial\rho_b}}=
\left(
\begin{array}{cc}
 \ums[3]{2}\la1 \varphi_1^2+\ums{2}\la{345}\varphi_2^2 & \la{345}\varphi_1\varphi_2 \\
 \la{345}\varphi_1\varphi_2 & \ums[3]{2}\la{2}\varphi_2^2+\ums{2}\la{345} \varphi_1^2 \\
\end{array}
\right)
+(\text{quadratic})\,.    
}
The term (quadratic) refers to the quadratic contribution given by
\eq{
(\text{quadratic})=\mtrx{m^2_{11} & -m^2_{12} \cr -m^2_{12} & m^2_{22}}\,.
}
The derivatives are with respect to the fields
\eq{
\phi_a=\mtrx{\phi_a^+\cr \displaystyle\frac{\rho_a+i\eta_a}{\sqrt{2}}}\,,\quad
a=1,2.
}
around the values in Eq.\,\eqref{vevs}.

\section{Cubic derivatives}
\label{ap:dddV}

The coefficients $\lambda_{i S}$ in \eqref{extremum:1-loop} are defined as
\eq{
\lambda_{i S}=\frac{1}{v_i}\frac{\partial M^2_{S}}{\partial v_i}
=\Big(U_{\rm sec}^\dag\frac{1}{v_a}\frac{\partial \cM_{\rm sec}}{\partial v_a}
U_{\rm sec}\Big)_{SS}\,,
}
where $S=\{G^+,H^+,G^0,A,H,h\}$, sec\,=\{char,\,odd,\,even\} refers to the different scalar sectors and $U_{\rm sec}$ diagonalizes $\cM_{\rm sec}$. The last subindex $SS$ refers to one of the diagonal entries following the ordering in \eqref{shifted.masses}.

Explicit calculation leads to 
\eqali{
\la{1G^+}&=\la1 c^2_{\beta_+}+\la3 s^2_{\beta_+}+\la{45}s_{\beta_+}c_{\beta_+}\frac{v_2}{v_1}\,,
\cr
\la{1H^+}&=\la1 s^2_{\beta_+}+\la3 c^2_{\beta_+}-\la{45}s_{\beta_+}c_{\beta_+}\frac{v_2}{v_1}\,,
\cr
\la{1G^0}&=\la1 c^2_{\beta_0}+\bla{345}s^2_{\beta_0}+2\la{5}s_{\beta_0}c_{\beta_0}\frac{v_2}{v_1}\,,
\cr
\la{1A^0}&=\la1 s^2_{\beta_0}+\bla{345}c^2_{\beta_0}-2\la{5}s_{\beta_0}c_{\beta_0}\frac{v_2}{v_1}\,,
\cr
\la{1H^0}&=3\la1 c^2_{\alpha'}+\la{345}s_{\alpha'}(s_{\alpha'}+2c_{\alpha'}\frac{v_2}{v_1})\,,
\cr
\la{1h^0}&=3\la1 s^2_{\alpha'}+\la{345}c_{\alpha'}(c_{\alpha'}-2s_{\alpha'}\frac{v_2}{v_1})\,,
}
where the mixing angles $\beta_+,\beta_0,\alpha'$ were defined in \eqref{shifted.angles}
with
\eqali{
\tan2\delta\beta_+&=\frac{(\dm[11]-\dm[22])s_{2\beta}}{m^2_{H^+}-(\dm[11]-\dm[22])c_{2\beta}}\,,
\cr
\tan2\delta\beta_0&=\frac{(\dm[11]-\dm[22])s_{2\beta}}{m^2_{A^0}-(\dm[11]-\dm[22])c_{2\beta}}\,,
\cr
\tan2\delta\alpha&=\frac{(\dm[11]-\dm[22])s_{2\alpha}}{-(m^2_{H^0}-m^2_{h^0})-(\dm[11]-\dm[22])c_{2\alpha}}\,.
}
The coefficients $\lambda_{2S}$ can be obtained from $\lambda_{1S}$ for each scalar $S$ above by using the replacement $Q_{12}$ in \eqref{Q12}.

\section{Cubic and quartic couplings}
\label{ap:cubics}

Given that the cubic and quartic couplings do not depend on the quadratic parameters,
they correspond to the tree-level ones listed, e.g., in Ref.\,\cite{gunion.haber} 
if we are in the limit $\alpha'\to \alpha$, $\beta_{0,+}\to \beta$.
If we want the couplings with these corrections, we can adapt the rotation angles $\alpha\to \alpha'$ or $\beta\to \beta_+$ or $\beta\to \beta_0$ for the couplings that do not mix the charged sector with the CP-odd sector because in the latter there is an ambiguity in distinguishing $\beta_+$ from $\beta_0$.
Another ambiguity arises if the couplings are written in terms of $(v,\beta)$ instead of $(v_1,v_2)$ because then we need to distinguish $\beta$ coming from the vevs and the $\beta_{0,+}$ coming from the diagonalization of the shifted mass matrices.

We adopt the convention that $-ig_{ABCD}$ is the Feynman rule associated to the 
vertex $ABCD$.
This is opposite to e.g.\ Ref.\,\cite{gunion.haber}.
We also abbreviate e.g.\ $g_{G^0G^0G^0G^0}$ as $g_{{G^0}^4}$.
The essential set of quartic couplings is
\subeqali{
\label{g:G0^4}
g_{{G^0}^4}
&=
3 \left[\la1 c^4_{\beta_0}+\la2 s^4_{\beta_0}+2 \la{345} 
s^2_{\beta_0} 
c^2_{\beta_0}\right]
\,,\\
\label{g:h^2G0^2}
g_{h^2{G^0}^2}
&=
\la1 s^2_{\alpha'} c^2_{\beta_0}
+\la2 c^2_{\alpha'} s^2_{\beta_0}
+\bla{345}c^2_{\beta_0-\alpha'}
-\ums{2}\la{345}s_{2 \alpha'} s_{2 \beta_0}
\,,\\
\label{g:HhG0^2}
g_{Hh{G^0}^2}
&=\ums{2}s_{2\alpha'}\big(-\la1 c^2_{\beta_0}+\la2 s^2_{\beta_0}
+\bla{345}c_{2\beta_0}\big)
+\la5 c_{2\alpha'}s_{2\beta_0}
\,,\\
\label{g:G0^2A^2}
g_{{G^0}^2{A}^2}
&=
\ums[3]{4}s^2_{2\beta_0}(\la1+\la2)+\ums{2}\la{345}c^2_{2\beta_0}
\,,\\
\label{g:G0^2HpHm}
g_{{G^0}^2H^+H^-}
&=
\la1 s^2_{\beta_+}c^2_{\beta_0}+\la2 s^2_{\beta_0}c^2_{\beta_+}
+\la3 c^2_{\beta_+-\beta_0}-\ums{2}\la{345}s_{2\beta_+}s_{2\beta_0}
\,,\\
\label{g:G0^2GpHm}
g_{{G^0}^2G^+H^-}
&=
\ums{2}s_{2\beta_+}(-\la1 c^2_{\beta_0}+\la2 s^2_{\beta_0}+\la3 c_{2\beta_0})
+\ums{2}\la{45}c_{2\beta_+}s_{2\beta_0}
\,,\\
\label{g:h^2AG0}
g_{h^2G^0A}
&=
\ums{2}s_{2 \beta_0}(-\la1  s^2_{\alpha'} +\la2  c^2_{\alpha'})
-\ums{2}\la{345}c_{2\alpha'}s_{2\beta_0}
+\ums{2}\la5 s_{2(\beta_0-\alpha')}
\,,\\
\label{g:A^3G0}
g_{G^0A^3}
&=
-\ums[3]{2}s_{2\beta_0}\Big[\la1  s^2_{\beta_0}-\la2 c^2_{\beta_0}
+\la{345}c_{2\beta_0}\Big]
\,,\\
\label{g:AG0HpHm}
g_{G^0AH^+H^-}
&=
-\ums{2}s_{2 \beta_0}\big(\la1 s^2_{\beta_+}-\la2  c^2_{\beta_+}
+\la3 c_{2 \beta_+}\big)
-\ums{2}\la{45}c_{2\beta_0}s_{2\beta_+} 
\,,\\
\label{g:AG0GpHm}
g_{G^0AG^+H^-}
&=
\ums{4}(\la1+\la2-2\la3)s_{2\beta_0}s_{2\beta_+}
+\ums{2}\la{45}c_{2\beta_0}c_{2\beta_+}
\,,\\
\label{g:HhAG0}
g_{HhG^0A}
&=
\ums{4}(\la1+\la2-2\bla{345})s_{2\beta_0}s_{2\alpha'}+\la5 c_{2\beta_0}c_{2\alpha'}
\,,\\
\label{g:hG0GpHm}
g_{hG^0H^\pm G^\mp}
&= -g_{HAH^\pm G^\mp} =
\pm\ums[i]{2}(-\la4+\la5)c_{\beta_0 - \alpha'}
\,.
}
More couplings can be obtained from simple replacements; see
table\,\ref{table.quartic}.
\begin{table}[h]
\[
\begin{array}{||c|c|c||}
\hline
 \text{Coupling}  & \text{Obtainable from}  & \text{Replacement}  \\
 \hline\hline
 g_{A^4}  & \eqref{g:G0^4}  & \beta_0\to \beta_0+\pi/2  \\
 g_{H^4}  & \eqref{g:G0^4}  & \beta_0\to \alpha'  \\
 g_{h^4}  & \eqref{g:G0^4}  & \beta_0\to \alpha'+\pi/2 \\
 g_{(G^+G^-)^2}  & \eqref{g:G0^4}  & \beta_0\to \beta_+ \\
 g_{(H^+H^-)^2}  & \eqref{g:G0^4}  & \beta_0\to \beta_+ +\pi/2 \\
 \hline
 g_{H^2{G^0}^2}  & \eqref{g:h^2G0^2}  & \alpha'\to\alpha'-\pi/2  \\
 g_{h^2{A}^2}  & \eqref{g:h^2G0^2}  & \beta_0\to\beta_0 +\pi/2  \\
 g_{H^2{A}^2}  & g_{h^2{A}^2} & \alpha'\to\alpha'-\pi/2 \\
 g_{Hh{A}^2}  & \eqref{g:HhG0^2}  & \beta_0\to\beta_0 + \pi/2 \\  
 g_{h^2{H}^2}  & \eqref{g:G0^2A^2}  & \beta_0\to\alpha'  \\
 g_{G^+G^-H^+H^-}  & \eqref{g:G0^2A^2}  & \beta_0\to\beta_+  \\
 g_{{A}^2H^+H^-}  & \eqref{g:G0^2HpHm}  & \beta_0\to \beta_0 +\pi/2  \\
 g_{{H}^2H^+H^-}  & \eqref{g:G0^2HpHm}  & \beta_0\to \alpha'  \\
 g_{{h}^2H^+H^-}  & \eqref{g:G0^2HpHm}  & \beta_0\to \alpha' + \pi/2  \\
 g_{{G^0}^2G^+G^-}  & \eqref{g:G0^2HpHm}  & \beta_+\to \beta_+ -\pi/2  \\
 g_{{A}^2G^+G^-}  & g_{{G^0}^2G^+G^-}  & \beta_0\to \beta_0 +\pi/2  \\
 g_{{H}^2G^+G^-}  & g_{{G^0}^2G^+G^-}  & \beta_0\to \alpha'  \\
 \hline 
\end{array}
\begin{array}{||c|c|c||}
\hline
 \text{Coupling}  & \text{Obtainable from}  & \text{Replacement}  \\
 \hline\hline
 g_{{h}^2G^+G^-}  & g_{{G^0}^2G^+G^-}  & \beta_0\to \alpha' + \pi/2  \\
 g_{{A}^2G^+H^-}  & \eqref{g:G0^2GpHm}  & \beta_0\to \beta_0 +\pi/2  \\
 g_{{H}^2G^+H^-}  & \eqref{g:G0^2GpHm}  & \beta_0\to \alpha'  \\
 g_{{h}^2G^+H^-}  & \eqref{g:G0^2GpHm}  & \beta_0\to \alpha' + \pi/2  \\
 g_{H^2G^0A}  & \eqref{g:h^2AG0} & \alpha'\to\alpha'-\pi/2  \\ 
 g_{h^3H} & \eqref{g:A^3G0}  & \beta_0\to \alpha'  \\ 
 g_{H^+H^-H^+G^-} & \eqref{g:A^3G0}  & \beta_0\to \beta_+  \\
 g_{{G^0}^3A} & (-1)\times\eqref{g:A^3G0}  & \beta_0\to \beta_0 +\pi/2  \\
 g_{H^3h} & (-1)\times\eqref{g:A^3G0}  & \beta_0\to \alpha' -\pi/2 \\
 g_{G^+G^-G^+H^-} & (-1)\times\eqref{g:A^3G0}  & \beta_0\to \beta_+ +\pi/2 \\
 g_{HhH^+H^-}  & \eqref{g:AG0HpHm}  & \beta_0\to \alpha'  \\
 g_{G^0AG^+G^-}  & \eqref{g:AG0HpHm}  & \beta_+\to \beta_+ -\pi/2  \\
 g_{HhG^+G^-}  & g_{G^0AG^+G^-}  & \beta_0\to \alpha'\\
 g_{HhG^+H^-}  & \eqref{g:AG0GpHm}  & \beta_0\to \alpha'  \\
 g_{G^+H^-G^+H^-}  & \eqref{g:HhAG0}  & \{\beta_0,\alpha'\}\to \beta_+ \\
 g_{HG^0H^\pm G^\mp}  & \eqref{g:hG0GpHm}  & \alpha'\to \alpha' - \pi/2 \\
 g_{hAH^\pm G^\mp}  & \eqref{g:hG0GpHm}  & \alpha'\to \alpha' - \pi/2 \\
 \hline
\end{array}
\]
\caption{\label{table.quartic}
Quartic couplings obtainable from reparametrization symmetry.}
\end{table}

The essential set of cubic couplings is
\subeqali{
\label{g:hG02}
g_{h{G^0}^2}
&=
-v_1\la1 c^2_{\beta_0}s_{\alpha'}+v_2\la2 s^2_{\beta_0}c_{\alpha'}
+\la5 s_{2\beta_0}(v_1 c_{\alpha'}-v_2s_{\alpha'})
+\bla{345}(v_2 c^2_{\beta_0}c_{\alpha'}-v_1s^2_{\beta_0}s_{\alpha'})
\,,\\
\label{g:hG0A}
g_{h{G^0}A}
&=
\ums{2}s_{2\beta_0}(v_1\la1 s_{\alpha'}+v_2\la2 c_{\alpha'})
+\la5 c_{2\beta_0}(v_1 c_{\alpha'}-v_2s_{\alpha'})
-\ums{2}\bla{345}s_{2\beta_0}(v_2c_{\alpha'}+v_1s_{\alpha'})
\,,\\
\label{g:hH2}
g_{hH^2}
&=
\ums[3]{2}s_{2\alpha'}(-v_1\la1 c_{\alpha'}+v_2\la2 s_{\alpha'})
+\ums{4}\la{345}[v_2(c_{\alpha'}+3c_{3\alpha'})-v_1(s_{\alpha'}-3s_{\alpha'})]
\,,\\
\label{g:hGpGm}
g_{hG^+G^-}
&=
v_2\la2 s^2_{\beta_+}c_{\alpha'}-v_1\la1 c^2_{\beta_+}s_{\alpha'}+\la3(v_2 c^2_{\beta_+}c_{\alpha'}-v_1 s^2_{\beta_+}s_{\alpha'})+\ums{2}\la{45}s_{2\beta_+}(v_1 c_{\alpha'}-v_2 s_{\alpha'})
\,,\\
\label{g:hGpHm}
g_{hG^+H^-}
&=
\ums{2}v_1(\la1-\la3)s_{2\beta_+}s_{\alpha'}+\ums{2}v_2(\la2-\la3) s_{2\beta_+}c_{\alpha'}+\ums{2}\la{45}c_{2\beta_+}(v_1 c_{\alpha'}-v_2 s_{\alpha'})
\,,\\
\label{g:G^0GpHm}
g_{G^0H^\pm G^\mp}
&=
\pm\ums[i]{2}(\la4 -\la5)(v_1 s_{\beta_0}-v_2 c_{\beta_0})
\,.
}
The remaining couplings can be obtained through reparametrization symmetries as shown in table\,\ref{table.cubic}.
\begin{table}[h]
\[
\begin{array}{||c|c|c||}
\hline
 \text{Coupling}  & \text{Obtainable from}  & \text{Replacement}  \\
 \hline\hline
 g_{H{G^0}^2}  & \eqref{g:hG02}  & \alpha'\to\alpha'-\pi/2  \\
 g_{h{A}^2}  & \eqref{g:hG02}  & \beta_0\to\beta_0 +\pi/2  \\
 g_{H{A}^2}  & g_{h{A}^2}  & \alpha'\to\alpha'-\pi/2  \\
 g_{h^3}  & 3\times\eqref{g:hG02}  & \beta_0\to\alpha'+\pi/2  \\
 g_{H^3}  & 3\times g_{H{G^0}^2}  & \beta_0\to\alpha'  \\
 g_{H{G^0}A}  & \eqref{g:hG0A} & \alpha'\to\alpha'-\pi/2  \\
\hline 
\end{array}
\begin{array}{||c|c|c||}
\hline
 \text{Coupling}  & \text{Obtainable from}  & \text{Replacement}  \\
 \hline\hline
 g_{Hh^2}  & \eqref{g:hH2}  & \alpha'\to\alpha'- \pi/2  \\
 g_{hH^+H^-} & \eqref{g:hGpGm}  & \beta_+\to\beta_+ + \pi/2 \\
 g_{HG^+G^-} & \eqref{g:hGpGm}  & \alpha'\to\alpha'- \pi/2 \\
 g_{HH^+H^-} & g_{HG^+G^-}  & \beta_+\to\beta_+ + \pi/2 \\
 g_{HG^+H^-} & \eqref{g:hGpHm}  & \alpha'\to\alpha'- \pi/2 \\
 g_{AH^\pm G^\mp} & \eqref{g:G^0GpHm}  & \beta_0\to\beta_0 + \pi/2 \\
 \hline
\end{array}
\]
\caption{\label{table.cubic}
Cubic couplings obtainable from reparametrization symmetry.}
\end{table}

Finally we note that, although the reparametrization symmetry already allows a huge simplification in the computation of the cubic and quartic couplings needed for our calculation, their use can be error-prone. In our routines we have adopted a different approach, namely we expanded the tree-level scalar potential in terms of physical fields $S=\{h,H,A,G^0,G^+,H^+,G^-,H^-\}$ and performed derivatives to obtain the desired couplings. For instance, 
\eqali{
g_{h^3G^0} = \frac{\partial^{4}V_{0}(S_i)}{\partial^{3} h\partial G^0}\Bigg|_{S_i=0},\quad
g_{hH^2} = \frac{\partial^{3}V_{0}(S_i)}{\partial^{2} h\partial H}\Bigg|_{S_i=0}
}

\section{Self-energy for the scalars}
\label{ap:self-energy}

We show here the different contributions to the self-energy of scalars due to scalars (S),
fermions (F) and gauge bosons (V) in the loop.

We first list the contributions from scalars in the loop for the different sectors.
For the CP odd sector we have
\subeqali{
16\pi^2\Pi^S_{G^0G^0}(s)&=
\ums{2}\sum_{S=G^0,A,H,h}g_{{G^0}^2S^2}A(M_S)
+\sum_{S=G^+,H^+}g_{{G^0}^2S\bar{S}}A(M_S)
\cr
&\quad
+\sum_{\stackrel{\mss{S=G^0,A}}{S'=H,h}}g_{G^0SS'}^2 B(M_S,M_{S'},s)
+2|g_{G^0G^+H^-}|^2 B(M_{G^+},M_{H^-},s)\,,
\\
16\pi^2\Pi^S_{AA}(s)&=
\ums{2}\sum_{S=G^0,A,H,h}g_{{A}^2S^2}A(M_S)
+\sum_{S=G^+,H^+}g_{{A}^2S\bar{S}}A(M_S)
\cr
&\quad
+\sum_{\stackrel{\mss{S=G^0,A}}{S'=H,h}}g_{ASS'}^2 B(M_S,M_{S'},s)
+2|g_{AG^+H^-}|^2 B(M_{G^+},M_{H^-},s)\,,
\\
16\pi^2\Pi^S_{AG^0}(s)&=16\pi^2\Pi^S_{G^0A}(s)
=
\ums{2}\sum_{S=G^0,A,H,h}g_{AG^0S^2}A(M_S)
+\sum_{S=G^+,H^+}g_{AG^0S\bar{S}}A(M_S)
\cr
&\quad
+\sum_{\stackrel{\mss{S=G^0,A}}{S'=H,h}}g_{ASS'}g_{G^0SS'} B(M_S,M_{S'},s)
+2g_{AG^+H^-}g^*_{G^0G^+H^-} B(M_{G^+},M_{H^-},s)\,.
\hs{3em}
}
Some couplings are absent because CP is conserved and $A,G^0$ are CP odd.

In the CP even sector we have
\subeqali{
16\pi^2\Pi^S_{HH}(s)&=
\ums{2}\sum_{S=G^0,A,H,h}g_{H^2S^2}A(M_S)
+\sum_{S=G^+,H^+}g_{H^2S\bar{S}}A(M_S)
\cr
&\quad
+\sum_{S,S'=G^0,A\text{~or~}H,h}g_{HSS'}^2 B(M_S,M_{S'},s)
+\sum_{S,S'=G^+,H^+}|g_{HS\bar{S}'}|^2 B(M_{S},M_{S'},s)
\,,
\hs{3em}
\\
16\pi^2\Pi^S_{hh}(s)&=
\ums{2}\sum_{S=G^0,A,H,h}g_{h^2S^2}A(M_S)
+\sum_{S=G^+,H^+}g_{h^2S\bar{S}}A(M_S)
\cr
&\quad
+\sum_{S,S'=G^0,A\text{~or~}H,h}g_{hSS'}^2 B(M_S,M_{S'},s)
+\sum_{S,S'=G^+,H^+}|g_{hS\bar{S}'}|^2 B(M_{S},M_{S'},s)
\,,
\hs{3em}
\\
16\pi^2\Pi^S_{Hh}(s)&=16\pi^2\Pi^S_{hH}(s)
=
\ums{2}\sum_{S=G^0,A,H,h}g_{HhS^2}A(M_S)
+\sum_{S=G^+,H^+}g_{HhS\bar{S}}A(M_S)
\cr
&\quad
+\sum_{S,S'=G^0,A\text{~or~}H,h}g_{HSS'}g_{hSS'} B(M_S,M_{S'},s)
+\sum_{S,S'=G^+,H^+}g_{HS\bar{S}'}g^*_{hS\bar{S}'} B(M_{S},M_{S'},s)
\,.
\cr
}

The self-energy for the charged sector is
\subeqali{
16\pi^2\Pi^S_{G^+G^+}(s)&=
\ums{2}\sum_{S=G^0,A,H,h}g_{G^+G^-S^2}A(M_S)
+\sum_{S=G^+,H^+}g_{G^+G^-S\bar{S}}A(M_S)
\cr
&\quad
+\sum_{\stackrel{\mss{S=G^+,H^+}}{S'=G^0,A,H,h}}|g_{G^+\bar{S}S'}|^2 B(M_S,M_{S'},s)
\\
16\pi^2\Pi^S_{H^+H^+}(s)&=
\ums{2}\sum_{S=G^0,A,H,h}g_{H^+H^-S^2}A(M_S)
+\sum_{S=G^+,H^+}g_{H^+H^-S\bar{S}}A(M_S)
\cr
&\quad
+\sum_{\stackrel{\mss{S=G^+,H^+}}{S'=G^0,A,H,h}}|g_{H^+\bar{S}S'}|^2 B(M_S,M_{S'},s)
\\
16\pi^2\Pi^S_{G^+H^+}(s)&=16\pi^2\Pi^S_{H^+G^+}(s)
=
\ums{2}\sum_{S=G^0,A,H,h}g_{G^+H^-S^2}A(M_S)
+\sum_{S=G^+,H^+}g_{G^+H^-S\bar{S}}A(M_S)
\cr
&\quad
+\sum_{\stackrel{\mss{S=G^+,H^+}}{S'=H,h}}g_{G^+\bar{S}S'}g^*_{H^+\bar{S}S'} B(M_S,M_{S'},s)
}
Note that couplings such as $g_{G^+G^-A^0}=0$ due to CP conservation.

The fermionic corrections to the propagator of a scalar $S$ to a scalar $S'$
are given by the general formula\,\cite{martin}:
\begin{align}
\Pi_{SS'}^F(s)=\frac{N_{c}}{16\pi^2}\Bigg\{
\frac{\tr[Y^{S}_{\bar{f}f'}(Y^{S'}_{\bar{f'}f})^{\dag}]}{2}B_{FF}(m_{f},m_{f'},s)
- m_{f}m_{f'}\tr[Y^{S}_{\bar{f}f'}\bar{Y}^{S'}_{\bar{f'}f}]B(m_{f},m_{f'},s)\Bigg\},
\end{align}
\noindent
where $N_{c}$ is the number of colors of the fermion in the loop, the $B$ function
was given in \eqref{B} while $B_{FF}(m_{f},m_{f'})$ is defined as
\begin{eqnarray}
B_{FF}(m_{f},m_{f'},s) &\equiv& [(s - m_{f} - m_{f'})B(m_{f},m_{f'},s) - A(m_{f}) - A(m_{f'})]
\,.
\end{eqnarray}
We denote the vertex of the scalar $S_{k}$ to the fermions $\bar{f}$ and $f'$ by $-iY^{k}_{\bar{f}f'}$
and they may contain the $\gamma_5$ matrix while $\bar{Y}$ refers to the transformation $\bar{\Gamma}=\gamma_0\Gamma^\dag\gamma_0$ in spinor space.
These Yukawa couplings are listed in table \ref{tab:yukawa} where the coefficients $C_{f}^{S}$ depend on the model used as in table \ref{tab:C}.

\begin{table}[h!]
\centering
\begin{tabular}{|c|c|c|c|}
\hline
        & $\bar{t}t$  & $\bar{b}b$ & $\bar{t}b$
\\ \hline
$h$   & $\frac{y_{t}}{\sqrt{2}}\frac{c_{\alpha'}}{s_{\beta}} \mathds{1}$  & $\frac{y_{b}}{\sqrt{2}}C^{\Mh}_{b}\mathds{1}$ & -
\\ \hline
$H$   & $\frac{y_{t}}{\sqrt{2}}\frac{s_{\alpha'}}{s_{\beta}}\mathds{1}$        & $\frac{y_{b}}{\sqrt{2}}C^{\MH}_{b}\mathds{1}$  & -
\\ \hline
$A$   & $-i\frac{y_{t}}{\sqrt{2}}\frac{c_{\beta_{0}}}{s_{\beta}} \gamma_{5}$ & $i\frac{y_{b}}{\sqrt{2}}C^{\MA}_{b}\gamma_{5}$  & -
\\ \hline
$G^0$   & $-i\frac{y_{t}}{\sqrt{2}}\frac{s_{\beta_{0}}}{s_{\beta}}\gamma_{5}$  & $i\frac{y_{b}}{\sqrt{2}}C^{\MG}_{b}\gamma_{5}$ &  -
\\ \hline
$H^{\pm}$ & - & - & $y_{b}C^{\MHpm}_{b}P_{R}-y_{t}\frac{c_{\beta_{+}}}{s_{\beta}}P_{L}$ 
\\ \hline
$G^{\pm}$ & - & - & $y_{b}C^{\MGpm}_{b}P_{R}-y_{t}\frac{s_{\beta_{+}}}{s_{\beta}}P_{L}$ 
\\ \hline
\end{tabular}
\caption{\label{tab:yukawa} Yukawa couplings}
\end{table}
\noindent

\begin{table}[h]
\centering
\begin{tabular}{|c|c|c|c|c|}
\hline
                & Type I                  & Type II                  \\ \hline
$C^{\Mh}_{b}$   & $c_\alpha'/s_\beta$     & $-s_\alpha'/c_\beta$     \\ \hline
$C^{\MH}_{b}$   & $s_\alpha'/s_\beta$     & $c_\alpha'/c_\beta$      \\ \hline
$C^{\MA}_{b}$   & $c_{\beta_{0}}/s_\beta$ & $-s_{\beta_{0}}/c_\beta$ \\ \hline
$C^{\MG}_{b}$   & $s_{\beta_{0}}/s_\beta$ & $c_{\beta_{0}}/c_\beta$  \\ \hline
$C^{\MHpm}_{b}$ & $c_{\beta_{+}}/s_\beta$ & $-s_{\beta_{+}}/c_\beta$ \\ \hline
$C^{\MGpm}_{b}$ & $s_{\beta_{+}}/s_\beta$ & $c_{\beta_{+}}/c_\beta$  \\ \hline
\end{tabular}
\caption{\label{tab:C}
Coefficients for type I and II 2HDM.
}
\end{table}

Finally, the contributions coming from the gauge bosons in the loop are given by
\begin{align}
\Pi_{SS'}^{V,\,\rm even}(s)=\frac{g^2}{16\pi^2}\Bigg\{&\frac{1}{4c_{w}^{2}} C_{SS''}^{eo}C_{S'S''}^{eo} B_{SV}(M_{S''},m_{Z},s)+\frac{1}{2}C_{SS''}^{e+}C_{S'S''}^{e+}B_{SV}(M_{S''},m_{W},s)\nonumber\\
&+\frac{m_{Z}^2}{2c_{w}^2}C_{S}^{0}C_{S'}^{0}B_{VV}(m_{Z},m_{Z},s)
+m_{W}^{2}C_{S}^{0}C_{S'}^{0}B_{VV}(m_{W},m_{W},s)\nonumber\\
&+\delta_{SS'}\left[\frac{3}{4c_{w}^2}A(m_{Z})+\frac{3}{2}A(m_{W})\right]
\Bigg\},
\end{align}
\begin{align}
\Pi_{SS'}^{V,\,\rm odd}(s)=\frac{g^2}{16\pi^2}\Bigg\{&\frac{1}{4c_{w}^{2}} C_{SS''}^{eo}C_{S'S''}^{eo} B_{SV}(M_{S''},m_{Z},s)+\frac{1}{2}C_{SS''}^{o+}C_{S'S''}^{o+}B_{SV}(M_{S''},m_{W},s)\nonumber\\
&+\delta_{SS'}\left[\frac{3}{4c_{w}^2}A(m_{Z})+\frac{3}{2}A(m_{W})\right]
\Bigg\},
\end{align}
\begin{align}
\Pi_{SS'}^{V,\,\rm char}(s)=\frac{g^2}{16\pi^2}\Bigg\{&\frac{1}{4}C_{SS''}^{e+}C_{S'S''}^{e+}B_{SV}(M_{S''},m_{W},s)+\frac{1}{4}C_{SS''}^{o+}C_{S'S''}^{o+}B_{SV}(M_{S''},m_{W},s)
\vspace{0.4cm} \nonumber \\
&+\delta_{SS'}s_{w}^{2}\left[ B_{SV}(M_{S},0,s)+ \cot_{2w}^2 B_{SV}(M_{S},m_{Z},s)\right]
\vspace{0.4cm} \nonumber \\
&+s_{w}^{2}C_{S}^{+}C_{S'}^{+}\left[s_{w}^{2}m_{Z}^2B_{VV}(m_{Z},m_{W},s)+m_{W}^{2}B_{VV}(0,m_{W},s)\right]
\vspace{0.4cm} \nonumber \\
&+\delta_{SS'}\left[3s_{w}^2\cot_{2w}^2 A(m_{Z})+\frac{3}{2}A(m_{W})\right]
\Bigg\},
\end{align}
\noindent
where there is an implicit summation on $S''$.
The loop functions $B_{SV}(m_{V},m_{S},s)$ and $B_{VV}(m_{V},m_{V'},s)$ are defined as
\begin{eqnarray}
B_{SV}(m_{S} ,m_{V},s) & = & \frac{(m_S^2 - m_{V}^2 - s)^2\,-\,4\,m_{V}^2\,s}{m_{V}^2}\,B(m_{S},m_{V},s) \,+\, A(m_{S}) \nonumber \\
& - &
\frac{(m_{S}^2 - s)^2}{m_{V}^2}\,B(m_{S},0,s) \,+\,
\frac{m_{S}^2 - m_{V}^2 - s}{m_{V}^2}\, A(m_{V}),
\label{eq:BSV}
\vspace{0.4cm}  \\
B_{VV}(m_{V},m_{V'},s) & = & 
2\,B(m_{V},m_{V'},s) \,+\,
\frac{(m_{V}^2 + m_{V'}^2 - s)^2}{4m_{V}^2 m_{V'}^2} \, B(m_{V},m_{V'},s)
\vspace{0.4cm} \nonumber \\
 & - &  \frac{(m_{V}^2 - s)^2}{4m_{V}^2 m_{V'}^2}\,B(m_{V},0,s) \,-\, \frac{(m_{V'}^2 - s)^2}{4m_{V}^2 m_{V'}^2}\,B(0,m_{V'},s)
\vspace{0.4cm} \nonumber \\
 & + &  \frac{p^4}{4m_{V}^2 m_{V'}^2} \, B(0,0,s) \,+\, \frac{A(m_{V})}{4\,m_{V}^2} \,+\,\frac{A(m_{V'})}{4\,m_{V'}^2}. 
 \label{eq:BVV}
\end{eqnarray}
\noindent
The couplings $C_{SS'}^{x}$ and $C_{S}^{x}$ can be read from the following tables.
\begin{table}[h!]
\centering
\begin{tabular}{|c|c|c|}
\hline
$C_{SS'}^{eo}$ & $\Mh$                   & $\MH$                    \\ \hline
$\MA$         & $c_{\beta_{0}-\alpha'}$ & $-s_{\beta_{0}-\alpha'}$ \\ \hline
$\MG$         & $s_{\beta_{0}-\alpha'}$ & $c_{\beta_{0}-\alpha'}$  \\ \hline
\end{tabular}
\begin{tabular}{|c|c|c|}
\hline
$C_{SS'}^{e+}$ & $\Mh$                   & $\MH$                    \\ \hline
$\MHpm$         & $c_{\beta_{+}-\alpha'}$ & $-s_{\beta_{+}-\alpha'}$ \\ \hline
$\MGpm$         & $s_{\beta_{+}-\alpha'}$ & $c_{\beta_{+}-\alpha'}$  \\ \hline
\end{tabular}
\begin{tabular}{|c|c|c|}
\hline
$C_{SS'}^{o+}$ & $\MA$                   & $\MG$                    \\ \hline
$\MHpm$         & $c_{\beta_{+}-\beta_{0}}$ & $-s_{\beta_{+}-\beta_{0}}$ \\ \hline
$\MGpm$         & $s_{\beta_{+}-\beta_{0}}$ & $c_{\beta_{+}-\beta_{0}}$  \\ \hline
\end{tabular}
\caption{\label{tab:C:SV}
Coefficients for the couplings between two scalars and one gauge boson.
}
\end{table}

\begin{table}[h!]
\begin{tabular}{|c|c|c|}
\hline
 & $\Mh$ & $\MH$ \\ \hline
$C_{S}^{0}$ & $s_{\beta-\alpha'}$ & $c_{\beta-\alpha'}$  \\
\hline
\end{tabular}
\begin{tabular}{|c|c|c|}
\hline
 & $\MHpm$ & $\MGpm$
\\ \hline
$C_{S}^{+}$ & $s_{\beta-\beta_{+}}$ & $c_{\beta-\beta_{+}}$
\\ \hline
\end{tabular}
\caption{\label{tab:C:VV}
Coefficients for the couplings between one scalar and two gauge bosons.
}
\end{table}

\section{Reparametrization symmetry}

Here we list some useful reparametrization symmetries that allow us to relate
different quartic and cubic scalar couplings that are numerous. These relations help us to check
different couplings or deduce new ones from a smaller set.
Moreover, given simple relations, they minimize errors and speed up numerical
implementation. 

The simplest reparametrization is to exchange fields of the same type, one in
$\phi_1$ and the other in $\phi_2$, together with a $90^\circ$ shift in the
respective mixing angle:
\eq{
Q_{\rm even}:\left\{\begin{aligned}
H &\to h \cr
h &\to -H \cr
\alpha' &\to \alpha'+\pi/2
\end{aligned}
\right.
,\quad
Q_{\rm odd}:\left\{\begin{aligned}
G^0 &\to A^0 \cr
A^0 &\to -G^0 \cr
\beta_0 &\to \beta_0+\pi/2
\end{aligned}
\right.
,\quad
Q_{\rm char}:\left\{\begin{aligned}
G^+ &\to H^+ \cr
H^+ &\to -G^+ \cr
\beta_+ &\to \beta_+ +\pi/2
\end{aligned}
\right.
.
}
This is what allowed us to relate $g_{H{G^0}^2}$ to $g_{h{G^0}^2}$ in
\eqref{g:hG02} after $\alpha'\to \alpha'-\pi/2$ since
$g_{h{G^0}^2}\cdot h{G^0}^2\to g_{h{G^0}^2}|_{\alpha'\to \alpha'-\pi/2}\cdot H{G^0}^2$ by the inverse transformation of $Q_{\rm even}$.
Another example is $g_{HhAG^0}$ in \eqref{g:HhAG0}: it is odd by the replacement
$\alpha'\to \alpha'+\pi/2$ or $\beta_0\to \beta_0+\pi/2$.

The reparametrization symmetry above arises by noting that the original fields in
$\phi_{1,2}$ are left invariant by the transformations and consequently the
potential is also invariant.
If we take the case of CP even fields,
\eq{
\mtrx{\rho_1\cr\rho_2}
=\mtrx{c_{\alpha'}H-s_{\alpha'}h\cr s_{\alpha'}H + c_{\alpha'}h}
\,,
}
we see $\rho_{1,2}$ are the same after the transformation $Q_{\rm even}$.
The same is true for the other pair of scalars and diagonalization angles.

The other reparametrization symmetry we can use is the exchange symmetry of the original potential \eqref{pot:tree}, restricted\,\footnote{%
Some adjustments are needed for the most general potential in \eqref{pot:tree}.} 
to the case of the CP conserving softly broken $\ZZ_2$,
\eq{
\phi_1\leftrightarrow\phi_2\,,\quad
m^2_{11}\leftrightarrow m^2_{22}\,,\quad
\la1\leftrightarrow\la2\,.
}
For the fields with definite masses at tree level this transformation reads
\eq{
\label{Q12}
Q_{12}:~~
\left\{
\begin{aligned}
m^2_{11}&\leftrightarrow m^2_{22}\,,\cr
\la1&\leftrightarrow\la2\,,\cr
v_1&\leftrightarrow v_2
\end{aligned}\right.
\left\{
\begin{aligned}
H&\leftrightarrow h\,,\cr
G^0&\leftrightarrow A\,,\cr
G^+&\leftrightarrow H^+\,,
\end{aligned}\right.
\left\{
\begin{aligned}
\alpha'&\to \pi/2-\alpha'\,,\cr
\beta_0&\to \pi/2-\beta_0\,,\cr
\beta_+&\to \pi/2-\beta_+\,.
\end{aligned}\right.
}

The last type of symmetry we can explore for reparametrization is the original gauge
invariance.
Discrete subgroups are the most useful.
For example, the reparametrization
\eq{
Q_{e\to o}:\quad \mtrx{H\cr h}\to \mtrx{G^0\cr A^0}\,,
\quad
\mtrx{G^0\cr A^0}\to -\mtrx{H\cr h}\,,
\quad
\left\{
\begin{aligned}
\alpha'&\leftrightarrow \beta_0\cr
v_1&\to -iv_1\cr
v_2&\to -iv_2
\end{aligned}\right.\,,
}
arises because of invariance by the gauge transformation
\eq{
\phi_a\to \mtrx{1&\cr&-i}\phi_a ,~~ a=1,2.
}

For special field configurations, we can also define an exchange reparametrization
between charged fields and CP even neutral fields as below
\eq{
Q_{c\leftrightarrow e}:\quad
\mtrx{H\cr h}\leftrightarrow \mtrx{G_1\cr H_1}\,,\quad
\alpha'\leftrightarrow \beta_+
,\qquad
\text{for $G^0=A^0=0$},
}
and $G^+=G_1/\sqrt{2}, H^+=H_1/\sqrt{2}$ can be chosen by electromagnetic gauge
invariance.
Similarly,
\eq{
Q_{c\leftrightarrow o}:\quad
\mtrx{G^0\cr A^0}\leftrightarrow \mtrx{G_2\cr H_2}\,,\quad
\beta_0 \leftrightarrow \beta_+
,\qquad
\text{for $H=h=0$},
}
and $G^+=iG_2/\sqrt{2}, H^+=iH_2/\sqrt{2}$ can be chosen.
These reparametrization symmetries arise from the gauge symmetry
\eq{
\phi_a\to \mtrx{0&1\cr1&0}\phi_a\,,\quad a=1,2.
}
Note that the vevs should also transform nontrivially and this reparametrization only works for quartic couplings.


\end{document}